\documentclass[12pt]{article}
\usepackage{amsmath}
\usepackage{amsfonts}
\usepackage{amssymb}
\usepackage{cite}
\def\hybrid{\topmargin -20pt    \oddsidemargin 0pt
        \headheight 0pt \headsep 0pt
        \textwidth 6.25in       
        \textheight 9.5in       
        \marginparwidth .875in
        \parskip 5pt plus 1pt   \jot = 1.5ex}
\hybrid
\numberwithin{equation}{section}
\numberwithin{table}{section}
\newcommand{\beq}{\begin{equation}}
\newcommand{\eeq}{\end{equation}}
\newcommand{\bi}{\begin{itemize}}
\newcommand{\ei}{\end{itemize}}
\newcommand{\bea}{\begin{eqnarray}}
\newcommand{\eea}{\end{eqnarray}}
\newcommand{\ba}{\begin{array}}
\newcommand{\ea}{\end{array}}
\newcommand{\bt}{\begin{tabular}}
\newcommand{\et}{\end{tabular}}
\newcommand{\bc}{\begin{center}}
\newcommand{\ec}{\end{center}}
\newcommand{\ax}{\alpha}
\newcommand{\bx}{\beta}
\newcommand{\cx}{\gamma}
\newcommand{\dx}{\delta}
\newcommand{\ox}{\omega}
\newcommand{\lx}{\lambda}
\newcommand{\ab}{\bar\alpha}
\newcommand{\bb}{\bar\beta}
\newcommand{\cb}{\bar\gamma}

\newcommand{\Ox}{\Omega}

\newcommand{\Gx}{\Gamma}
\newcommand{\cC}{\mathcal{C}}

\newcommand{\cL}{\mathcal{L}}
\newcommand{\cK}{\mathcal{K}}
\newcommand{\cN}{\mathcal{N}}

\newcommand{\cF}{\mathcal{F}}

\newcommand{\cM}{\mathcal M}
\newcommand{\W}{\mathcal W}

\newcommand{\wg}{\wedge}

\newcommand{\del}{\partial}

\DeclareMathOperator{\SU}{\mathit{SU}}
\DeclareMathOperator{\SO}{\mathit{SO}}

\DeclareMathOperator{\Spin}{\mathit{Spin}}
\DeclareMathOperator{\so}{\mathit{so}}
\DeclareMathOperator{\su}{\mathit{su}}

\newcommand{\rep}[1]{\mathbf{#1}}

\newcommand{\dd}{\mathrm{d}}
\newcommand{\ee}{\mathrm{e}}
\newcommand{\ii}{\mathrm{i}}

\newcommand{\bbZ}{\mathbb{Z}}
\newcommand{\bbR}{\mathbb{R}}
\newcommand{\bbC}{\mathbb{C}}

\newcommand{\CY}{Calabi--Yau}

\newcommand{\nn}{\nonumber}

\newcommand{\tox}{\tilde\omega}
\newcommand{\txi}{\tilde\xi}
\newcommand{\IM}{\textrm{Im} \,}
\newcommand{\RE}{\textrm{Re} \,}

\newcommand{\p}{\tilde m}
\newcommand{\q}{\tilde e}
\newcommand{\ff}{{\zeta}}
\newcommand{\CT}{\kappa}          
\newcommand{\LCY}{\tilde{L}}      

\begin{document}

\begin{titlepage}
\begin{center}

\hfill QMW-PH-02-21\\
\hfill hep-th/0211102\\

\vfill
{\large \bf  Mirror Symmetry in Generalized Calabi--Yau 
  Compactifications}\footnote{%
Work supported by: DFG -- The German Science Foundation,
GIF -- the German--Israeli Foundation for Scientific Research,
European RTN Program HPRN-CT-2000-00148 and the
DAAD -- the German Academic Exchange Service, 
the CNRS -- the French Center for National Scientific Research 
and the Royal Society, UK.}\\

\vskip 0.8cm

{\bf Sebastien Gurrieri}\\

\vskip 0.3cm
{\em Centre de Physique Th\'eorique, CNRS Luminy, Case 907,\\
F-13288 Marseille -Cedex 9, France}\\
{\tt gurrieri@cpt.univ-mrs.fr}
\vskip 0.8cm
{\bf Jan Louis,  Andrei Micu\footnote{On leave from IFIN-HH Bucharest}
}  \\

\vskip 0.3cm
{\em Fachbereich Physik, Martin-Luther-Universit\"at Halle-Wittenberg,\\
Friedemann-Bach-Platz 6, D-06099 Halle, Germany}\\
{\tt j.louis@physik.uni-halle.de, micu@physik.uni-halle.de}

\vskip 0.5cm

{\bf Daniel Waldram}\\

\vskip 0.3cm
{\em Department of Physics, Queen Mary, University of London, \\
Mile End Road, London E1 4NS, United Kingdom}\\
{\tt d.j.waldram@qmul.ac.uk}

\end{center}

\vskip 1cm

\begin{center} {\bf ABSTRACT } \end{center}

\noindent
We discuss mirror symmetry in generalized Calabi--Yau
compactifications of type~II string theories
with background NS fluxes.
Starting from type IIB compactified on Calabi--Yau
threefolds with NS three-form flux we show 
that the mirror type IIA theory arises from
a purely geometrical compactification on
a different class of six-manifolds.
These mirror manifolds have $SU(3)$ structure and are termed 
{\it half-flat}; they are neither complex nor Ricci-flat and their
holonomy group is no longer $\SU(3)$. 
We show that type IIA appropriately compactified on such manifolds
gives the correct mirror-symmetric low-energy effective action. 

\vfill

\noindent November 2002

\end{titlepage}

\section{Introduction}
\setcounter{equation}{0}

In ten space-time dimensions ($D=10$)
there exist two inequivalent type II string theories denoted type IIA
and type IIB. Both theories have the maximal amount
of 32 local supersymmetries but they differ 
in their field content~\cite{GSW,LT,JP}.
{}From a phenomenological point of view it is of interest
to study their compactifications with less supersymmetry 
and a space-time background of the form $\bbR^{1,3}\times Y$. 
Here $\bbR^{1,3}$ denotes four-dimensional  Minkowski space
while $Y$ is a compact six-dimensional Euclidean manifold whose
holonomy group determines the amount of supersymmetry which
is left unbroken by the background. 
If the holonomy group is trivial all 32 supercharges are preserved
while $\SO(6)$ (or any subgroup thereof) breaks all (or some) of the
supercharges. 
Calabi--Yau threefolds are a particularly interesting class of
compactification manifolds as their holonomy group
is $SU(3)$ and as a consequence
they only leave eight supercharges intact~\cite{GSW,LT,JP}.

In a compactification on a Calabi--Yau threefold, the light modes of
the effective theory all appear as form-field zero modes of the
Laplace operator on $Y$. Such harmonic forms are in
one-to-one correspondence with non-trivial elements of the cohomology
groups $H^{(p,q)}(Y)$. The interactions of the light modes are
captured by a low-energy effective Lagrangian ${\cal L}_{{\rm eff}}$
which can be computed via a Kaluza-Klein (KK) reduction of the ten-dimensional
Lagrangian. This low-energy theory is found to be a four-dimensional
$N=2$ supergravity coupled to vector-, tensor- and
hypermultiplets~\cite{wp,N=2,BCF,BGHL}.  

The low-energy effective theories of type IIA and type IIB 
in $D=4$ are not unrelated. Mirror symmetry assembles topologically
distinct Calabi--Yau threefolds into `mirror pairs'  
$(Y,\tilde Y)$ such that type IIA compactified on $Y$ is
equivalent to type IIB compactified on the mirror manifold $\tilde
Y$~\cite{mirror}. This leads to a relation between even (or odd)
cohomology groups on $Y$ and odd (or even) cohomology groups on
$\tilde{Y}$. Thus, for instance, the Euler numbers $\chi$ of the
mirror pair have opposite signs  $\chi(Y)=-\chi(\tilde Y)$.  

Among their massless excitations both type II string theories contain
$(p-1)$-form gauge potentials $C_{p-1}$ with a $p$-form field strength
$F_{p} = \dd C_{p-1}$. Recently generalized Calabi--Yau compactifications
of type II string theories have been considered where background
fluxes for the field strengths $F_{p}$ on $Y$ are turned 
on~\cite{PS,JM,TV,PM1,CKLT,GKP,CKKL,GD,LM2,CKL}.\footnote{%
The closely related heterotic and orientifold compactifications are discussed,
for example, in
refs.~\cite{AS1,Hull1,dWS,Bachas,DRS,LM1,KST,AAHV,BBHL,BBKL,FP,BD}.} 
More precisely, one allows $F_p$ of the form
\begin{equation}
F_{p} = e_i \omega_{p}^i\ ,
\end{equation}
where $e_i\omega_p^i$ is a general harmonic $p$-form on $Y$, written in
terms of a harmonic basis $\omega_p^i$ of the group $H^p(Y,\bbR)$. The
harmonic condition is required to ensure that the Bianchi identity and
the equation of motion are left intact so 
\begin{equation}
 \dd F_p =0= \dd^\dagger F_p \ .
\end{equation}
Note that this implies that the gauge potential $C_{p-1}$ is only
locally defined on $Y$. 
Integrating $F_{p}$ over the $p$-cycle $\gamma_p^i$ in $Y$ which is
Poincar\'{e}-dual to $\omega_p^i$ gives 
\begin{equation}\label{fluxdef}
  \int_{\gamma_p^i} F_p  = e_i \ .
\end{equation}
Due to a Dirac quantization condition, the flux $e_i\omega^i$ is
quantized in string theory meaning it is actually an element of
integral cohomology $H^p(Y,\bbZ)$. Choosing a basis $\omega^i_p$
which is also a basis of $H^p(Y,\bbZ)$ (we ignore here any torsion
elements in the integral cohomology) this means that the flux
parameters $e_i$ are integers. The number of parameters is simply
given by the Betti number which is the dimension of the appropriate
cohomology group $H^p(Y,\bbR)$. On a Calabi--Yau manifold, the
only odd cohomology group is $H^3(Y)$, while all the even
cohomology groups $H^0(Y)$, $H^2(Y)$, $H^4(Y)$ and $H^6(Y)$ are
present.  

The flux parameters contribute to the energy-momentum tensor and as a
consequence
the geometry backreacts and a non-trivial warp-factor is induced 
\cite{dWS,AS1,Hull1}.
However, in the supergravity approximation
the cycles $\gamma_p^i$ are chosen to be large and hence
the fluxes parameters are effectively
continuous and  represent small perturbations
of the original Calabi--Yau compactification. This in turn implies
that the light modes are still determined by the linear
fluctuations around the background values (zero modes) of the theory in the 
absence
of fluxes. In this approximation their induced masses are much smaller
than the integrated out heavy states with masses of order the string scale
or the compactification scale.
Thus the interactions of the light modes continue
to be captured
by an effective Lagrangian ${\cal L}_{{\rm eff}}$ which now 
depends continuously
on the flux parameters $e_i$. The fluxes appear
as gauge or mass parameters and deform the original supergravity
into a gauged or massive supergravity.
The fluxes introduce a non-trivial potential for some
of the massless fields and spontaneously break (part of) the supersymmetry.

${\cal L}_{{\rm eff}}$ has been computed in various situations.
In refs.~\cite{JM,TV,GD,LM2} type IIB compactified
on Calabi--Yau threefolds $\tilde{Y}$ in the presence of
RR-three-form flux $F_3$ and NS-three-form flux $H_3$
was derived.
In refs.~\cite{PS,CKKL,LM2} type IIA compactified on
the mirror manifold $Y$ with RR-fluxes
$F_0$, $F_2$, $F_4$ and $F_6$ present was considered.
The resulting low-energy effective action was equivalent 
to the type IIB action on the mirror manifold $\tilde{Y}$ with $F_3$
non-zero, but $H_3=0$~\cite{LM2}. As expected, given the matching of
odd and even cohomologies on mirror pairs, the type IIB RR-fluxes $F_3$
in the third cohomology group $H^3(\tilde{Y})$ are mapped to the type IIA
RR-fluxes in the even cohomology groups $H^0(Y)$, $H^2(Y)$, $H^4(Y)$
and $H^6(Y)$~\cite{GVW,gukov}. 

However, for non-vanishing NS-fluxes the situation is less clear as
no obvious mirror symmetric compactification is known.
In both type IIA and type IIB on $Y$ an NS three-form $H_3$ exists
which can give a non-trivial NS-flux in $H^3(Y)$. However, in neither
case are there NS form fields which can give fluxes in the mirror
symmetric even cohomologies $H^0(Y)$, $H^2(Y)$, $H^4(Y)$ and
$H^6(Y)$. Vafa~\cite{Vafa} suggested that the mirror symmetric
configuration is related to compactifying on a manifold $\hat{Y}$
which is not complex but only admits an almost complex structure whose
Nijenhuis tensor is non-vanishing. The purpose of this paper is to
make this proposal more precise. 

As a first step we demand that the $D=4$ effective action continues to have
$N=2$ supersymmetry, that is, eight local supersymmetries. This
implies that there is a single globally defined spinor $\eta$ on $\hat
Y$ so that each of the $D=10$ supersymmetry parameters gives a single
local four-dimensional supersymmetry. As result, the structure group
of the bundle of orthonormal frames on $\hat Y$ has to reduce from
$\SO(6)$ to $SU(3)$. If we further demand that the two $D=4$
supersymmetries are unbroken in a Minkowskian ground state $\eta$ has
to be covariantly constant with respect to the Levi-Civita connection
$\nabla$ or equivalently the holonomy group has to be $SU(3)$. This
second  requirement uniquely singles out Calabi--Yau threefolds as the
correct compactification manifolds.
However, in this paper we relax this second condition
and only insist that a globally defined $\SU(3)$-invariant spinor
exists. Manifolds with this property have been discussed in the mathematics
and physics literature and are known as manifolds with $SU(3)$ structure
(see, for example, refs.~\cite{FFS,salamonb,joyce,friedrich,salamon,
CS,AS1,Hull1,rocek,papa,waldram,KMPT}). 
They admit an almost complex structure $J$, a metric $g$ which is
hermitian with respect to $J$ and a unique $(3,0)$-form $\Omega$. 
Generically, since $\eta$ is no longer covariantly constant, the
Levi-Civita connection now fails to have $\SU(3)$-holonomy. However
one can always write $\nabla\eta$ in terms of a three-index tensor,
$T^0$, contracted with gamma matrices, acting on $\eta$. In the same
way $\nabla J$ and $\nabla\Omega$ can be also written in terms of
contractions of $T^0$ with $J$ and $\Omega$ respectively. This tensor
$T^0$, known as the intrinsic torsion, is thus a measure of the
obstruction to having $\SU(3)$ holonomy.

Different classes of manifolds with $\SU(3)$ structure exist and they
are classified by the different elements in the decomposition of the
intrinsic torsion into irreducible $\SU(3)$ representations. 
We will mostly consider the slightly non-generic situation where only
``electric'' flux is present. 
In this case, we find that mirror symmetry restricts us to a
particular class of manifolds with $SU(3)$ structure called 
{\it half-flat} manifolds \cite{CS}.\footnote{%
Manifolds with torsion have also been considered  
in refs.~\cite{AS1,Hull1,BD,friedrich,rocek,papa,waldram}.
However, in these papers the torsion
is usually chosen to be completely antisymmetric in its indices
or in other words it is a three-form. This turns out
to be a different condition on the torsion and these manifolds
are not half-flat.}
They are neither complex, nor K\"ahler, nor Ricci-flat but they are
characterized by the conditions
\begin{equation}
\dd \Omega^- =\ 0 =\ \dd (J\wedge J) \ ,
\end{equation}
where $\Omega^-$ is the imaginary part of the $(3,0)$-form.
On the other hand the real part of $\Omega$ is not closed
and plays precisely the role of an NS four-form
$d\Omega^+ \sim F_4^{NS}$ corresponding to fluxes along
$H^4(Y)$~\cite{Vafa}. Thus the `missing' NS-fluxes are purely
geometrical and arise directly from  the change in the
compactification geometry. 

Half-flat manifolds also appear from a different point of view.
When appropriately fibered over an interval the resulting seven-dimensional
manifold always admits a metric of $G_2$ holonomy~\cite{CS,H}.
Physically this corresponds to the fact that the effective
four-dimensional $N=2$ theory has $N=1$ BPS domain-wall (DW)
solutions which are mirror symmetric to the type IIB
DW solutions studied in \cite{BCL}. In fact, all these
DW solutions are exact solutions of the full ten-dimensional
supergravity theory without any need to assume the relevant NS flux is
small. Typically one can only expect the KK-reduction to
be consistent in the limit where the flux is a small perturbation.  
This picture is also related to work in~\cite{AMV}. 
There it was shown that starting from a type IIB theory with both RR-
and NS background fluxes the conjectured mirror symmetric type IIA
theory is related to a purely geometrical compactification of M-theory
on a $G_2$ manifold. 

This paper is organized as follows. In section~\ref{mirrorreview}
we briefly recall mirror symmetry in Calabi--Yau compactifications with
RR-flux. In section~\ref{Yhat} we discuss 
properties of manifolds with $SU(3)$ structure and the way they
realize supersymmetry in the effective action.
These manifolds are classified in terms of irreducible
representations of the structure group $SU(3)$ and in section~\ref{sec:hf}
we argue that the class of half-flat manifolds are likely
to be the mirror geometry of Calabi--Yau manifolds with
electric NS-fluxes.
We confirm this `educated guess' in section~\ref{sec:torus}
by considering a complex six-torus (and implicitly orbifolds
thereof~\cite{VW}) where mirror symmetry is directly related to
T-duality and thus the mirror manifold can be explicitly
constructed.\footnote{%
This has also been considered recently in refs.~\cite{dh,kachru} and
our discussion in section 2.4 overlaps with these papers.
We thank the authors of \cite{kachru} for communicating their results
prior to publication. T-duality in massive supergravities
has been discussed in refs.~\cite{hull,KM,HLS,janssen,BBRS,singh}.}
This can be slightly generalized by considering Calabi--Yau manifolds
in the SYZ picture where $Y$ is a special Lagrangian $T^3$
fibration~\cite{SYZ}. In this case mirror symmetry is also related to
T-duality and, in the large complex structure 
limit, can be carried out explicitly. 
In both cases we discover that half-flat manifolds arise as
the mirror symmetric geometry. This is further confirmed
in section~\ref{sec:DM} where we discuss $N=1$ BPS-domain wall solutions
and their relation to manifolds with $SU(3)$ structure and 
manifolds with $G_2$ holonomy.
In section~\ref{KK} we perform the KK-reduction of
type IIA compactified on $\hat Y$,
derive the low-energy effective action and show that it is mirror
symmetric to type IIB compactified on threefolds $Y$ with non-trivial
electric NS-flux $H_3$. The effect of the altered geometry
is as expected.
It turns an ordinary supergravity into a gauged supergravity
in that scalar fields become charged and a potential is induced.
This potential receives contributions from different terms in
ten-dimensional effective action, one of which arises from the 
non-vanishing Ricci-scalar. This contribution is crucial to obtain the 
exact mirror symmetric form of the potential.

The derivation of the
mirror symmetric effective action including
magnetic fluxes is technically more involved due to the appearance of a
massive RR two-form. This in turn requires a KK-reduction
on the `democratic' version of the ten-dimensional type IIA
effective action \cite{BKORV,GH}
and we postpone this study to a separate publication \cite{GLMW}.
Section \ref{conc} contains our conclusions.
Some of the technical details are relegated to four appendices.
In appendix \ref{conv} we summarize our conventions. In order to make the
paper self-contained we recall in appendix \ref{IIAcomp}
the Calabi--Yau compactification of type IIA without fluxes
while in appendix \ref{IIBNSflux} we recall the effective action for
type IIB compactification on Calabi--Yau manifolds with
non-trivial NS-flux.
In appendix \ref{acs}
we discuss manifolds with $G$-structure from a more
mathematical point of view and supply some explicit expressions
omitted in the main text.
In appendix \ref{Rhf} we compute the Ricci-scalar for half-flat manifolds,
show that it is non-zero and hence contributes to the scalar potential.


\section{Generalized mirror manifolds}


\subsection{Mirror symmetry in Calabi--Yau compactifications with flux}
\label{mirrorreview}
Let us begin by reviewing mirror symmetry for Calabi--Yau
compactifications with non-trivial fluxes. Recall that the two
ten-dimensional type II theories, compactified on a Calabi--Yau
manifold $Y$, each lead to a four-dimensional low-energy effective action
which is an $N=2$ supergravity coupled to vector-, 
tensor- and hypermultiplets \cite{wp,N=2,BCF,BGHL}. More
precisely, for type IIA the massless spectrum contains $h^{(1,1)}$
vector multiplets, $h^{(1,2)}$ hypermultiplets and one tensor
multiplet while for type IIB one has $h^{(1,2)}$ vector multiplets,
$h^{(1,1)}$ hypermultiplets and one tensor multiplet. Here,
the Hodge numbers $h^{(1,1)}$ and $h^{(1,2)}$ are the dimensions
of the cohomology groups $H^{1,1}(Y)$ and $H^{1,2}(Y)$.
In appendix~\ref{IIBnsflux} we review some of the details of these
compactifications and give explicitly the effective
action. 

Calabi--Yau manifolds $Y$ and $\tilde{Y}$ form a mirror pair
if compactifying type IIA on $Y$ gives the same theory as
compactifying type IIB on $\tilde Y$ \cite{mirror}. 
More precisely one requires that
the corresponding string superconformal field theories including
quantum corrections are equivalent. This implies, among
other things, that they have reversed Hodge numbers,
\begin{equation}
  h^{(1,1)}(\tilde Y)= h^{(1,2)}(Y)\ , \qquad
  h^{(1,2)}(\tilde Y) = h^{(1,1)}(Y)\ ,
\end{equation}
and, that the two effective $D=4$ Lagrangians coincide
\begin{equation}
{\cal L}^{\text{(IIA)}}(Y)\ \equiv\
   {\cal L}^{\text{(IIB)}}(\tilde Y)\ .
\end{equation}

In the supergravity limit this symmetry continues to hold when
background RR-flux is included on the Calabi--Yau manifolds. Consider
first type IIB. The only allowed RR-flux on the internal Calabi--Yau
manifold $\tilde{Y}$ is the three-form $F_3=dC_2$. It must be harmonic
and so is parameterized by an element of the cohomology group
$H^3(\tilde{Y},\bbR)$. In string theory, the flux is quantized so is
more correctly an element of the integer cohomology
$H^3(\tilde{Y},\bbZ)$. This allows the possibility that $F_3$ includes
``torsion'' elements, that is non-zero elements of
$H^3(\tilde{Y},\bbZ)$ the image of which in $H^3(\tilde Y,\bbR)$
vanishes. (This should not be confused with the notion of torsion of a
metric-compatible connection which will be central to later
discussions in this section.) Here we will ignore such subtleties and
assume such elements vanish.\footnote{It has been argued that the 
  flux is really described by an element of K-theory~\cite{MW}. This
  differs from $H^3(\tilde{Y},\bbZ)$ precisely in the subgroup of
  torsion elements, hence again here we will ignore this distinction.}
In general, one can introduce a symplectic integral basis $\ax_A$,
$\bx^A$, with $A=0,\dots,h^{(1,2)}$ for $H^3(\tilde Y,\bbR)$. The flux
$F_3$ then defines $2(h^{(1,2)}+1)$ flux parameters $(\q_A, \p^A)$
according to
\begin{equation}
F_3 = \dd C_2 + \p^A \ax_A + \q_A \bx^A   \ .
\end{equation}
The effective action of this compactification is worked out via a
Kaluza--Klein reduction in refs.~\cite{JM,TV,GD,LM2}. It uses
the supergravity limit where the flux parameters are small compared to the
string scale and the backreaction of the Calabi--Yau geometry to the
presence of the fluxes is assumed to excite only the zero modes of the
Calabi--Yau manifold. In other words,  
a KK reduction is performed on the original Calabi--Yau geometry albeit
  with the non-vanishing fluxes taken into account. 
This leads to a potential which induces perturbatively small masses
for some of the scalar fields and spontaneously breaks supersymmetry.

It was shown in~\cite{LM2} that this IIB effective action is
manifestly mirror symmetric to the one arising from the
compactification of massive type IIA supergravity \cite{Romans} on $Y$
with RR-fluxes turned on in the even cohomology of $Y$. More
precisely, in IIA compactifications the RR two-form field strength
$F_2$ can have non-trivial flux in $H^2(Y,\bbZ)$ while the four-form
field strength $F_4$ has fluxes in $H^4(Y,\bbZ)$. (Again we
ignore any torsion elements.) Let $\omega_i$ with $i=1,\ldots,h^{(1,1)}$
be an integral basis of $H^2(Y,\bbR)$ and $\tilde\omega^i$ be an
integral basis of $H^4(Y,\bbR)$. Then there are $2h^{(1,1)}$ IIA RR-flux
parameters given
by
\begin{equation}
F_2 = \dd A_1 +  m^i \omega_i \ , \qquad
F_4 = \dd C_3 -A_1\wedge H_3 + e_i \tilde\omega^i  \ .
\end{equation}
In addition there are the two extra parameters $m^0$ and $e_0$,
where $e_0$ is the dual of the space-time part of the four-form
$F_{4\,\mu\nu\rho\sigma}$ and $m^0$ is the mass parameter of the
original ten-dimensional massive type IIA theory \cite{LM2}.
Altogether there are $2(h^{(1,1)}+1)$ real RR-flux parameters
$(e_I,m^J), I,J=0,1,\ldots, h^{(1,1)}$  which precisely map to the
$2(h^{(1,2)}+1)$ type IIB RR-flux parameters under mirror symmetry. This
is confirmed by an explicit KK-reduction of the respective effective
actions and one finds~\cite{LM2}\footnote{For $m^I = 0$ one finds a
  standard $N = 2$ gauged supergravity with a potential for the moduli
  scalars of the vector multiplets. For $m^I \neq 0$ a non-standard
  supergravity occurs where the two-form $B_2$ becomes massive. For a more
  detailed discussion and a derivation of the effective action we refer
  the reader to ref.\ \cite{LM2}.}
\begin{equation}
{\cal L}^{(IIA)}(Y, e_I, m^J)\ \equiv\ {\cal L}^{(IIB)}(\tilde Y,\q_A, \p^B)\ .
\end{equation}

We expect that mirror symmetry continues to hold when one considers
fluxes in the NS-sector. However, in this case, the situation is more
complicated. In both type IIA and type IIB there is a NS two-form
$B_2$ with a three-form field strength $H_3$, so one can consider
fluxes in $H^3(Y,\bbZ)$ in IIA and $H^3(\tilde{Y},\bbZ)$ in
IIB. However, these are clearly not mirror symmetric backgrounds since
mirror symmetry exchanges the even and odd cohomologies. One appears
to be missing $2(h^{(1,1)}+1)$ NS-fluxes, lying along the even
cohomology of $Y$ and $\tilde Y$, respectively.
Since the NS fields include only the metric,
dilaton and two-form $B_2$, there is no candidate NS even-degree
form-field strength to provide the missing fluxes. Instead, they must
be generated by the metric and the dilaton. Thus we are led to
consider compactifications on a generalized class of manifolds
$\hat{Y}$ with a metric which is no longer Calabi--Yau, and perhaps
a non-trivial dilaton in order to find a mirror-symmetric effective
action. This necessity was anticipated by Vafa in
ref.~\cite{Vafa}.

We now turn to what characterizes this generalized class of
compactifications on $\hat{Y}$.  For definiteness, we will pose the
problem as one of finding the IIA compactifications on $\hat{Y}$
mirror to IIB compactifications on the Calabi--Yau manifold
$\tilde{Y}$ with NS flux $H_3$. Since the NS sectors of IIA and IIB are
identical this 
is, of course, equivalent to the problem with the roles of IIA and IIB
reversed.  The low-energy effective action of the IIB
theory with NS flux 
\begin{equation}
\label{H3flux}
   H_3 = \dd B_2 + \p^A \ax_A + \q_A \bx^A   \ .
\end{equation}
can be easily calculated as is done in appendix~\ref{IIBNSflux}. 
Following the usual convention, we refer to $\q_A$ and $\p^A$ as
electric and magnetic fluxes respectively. We will generally consider the
IIA dual of the pure electric case where only half the fluxes
in~\eqref{H3flux} are excited.


\subsection{Supersymmetry and manifolds with $\SU(3)$-structure}

\label{Yhat}

The low-energy effective action arising from IIB compactifications
with non-trivial $H_3$-flux 
describes a massive deformation of an $N=2$ supergravity \cite{JM,TV,GD,LM2}. 
Compactification on the
conjectured generalized mirror IIA manifold $\hat{Y}$ should lead to
the same effective action. Thus the first constraint on $\hat{Y}$ is
that the resulting low-energy theory preserves $N=2$
supersymmetry.

Let us first briefly review how supersymmetry is realized in the
conventional Calabi--Yau compactification. Ten-dimensional type IIA
supergravity has two supersymmetry parameters $\epsilon^\pm$ of
opposite chirality each transforming in a real 16-dimensional spinor
representation of the Lorentz group $\Spin(1,9)$. In particular, the
variation of the two gravitinos in type IIA is schematically given
by~\cite{GP} 
\begin{equation}\label{gravitino}
\delta \psi_M^\pm =
   \left[\nabla_M + (\Gamma \cdot H_3)_M\right]\epsilon^\pm
   + \left[(\Gamma \cdot F_2)_M
      + (\Gamma \cdot F_4)_M\right]\epsilon^\mp + \ldots \ ,
\end{equation}
where the dots indicate further fermionic terms. 
Next one dimensionally
reduces on a six-dimensional manifold $Y$ and requires that the theory
has a supersymmetric vacuum of the form $\bbR^{1,3}\times Y$ with all
other fields trivial.  This implies that there are
particular spinors $\epsilon^\pm$ for which the gravitino
variations~\eqref{gravitino} with $H_3=F_2=F_4=0$ vanish. On 
$\bbR^{1,3}\times Y$ the Lorentz group $\Spin(1,9)$ decomposes into
$\Spin(1,3)\times\Spin(6)$ and we can correspondingly write
$\epsilon^\pm=\theta^\pm\otimes\eta$. In the supersymmetric vacuum,
the vanishing of the gravitino variations imply the $\theta^\pm$ are 
constant and $\eta$ is a solution of   
\begin{equation}
\label{ccons}
\nabla_m \eta=0\ , \qquad m = 1,\dots,6 \ .
\end{equation}
If this equation has a single solution, each $\epsilon^\pm$ gives a
Killing spinor and we see that the background preserves $N=2$
supersymmetry in four dimensions as required. Equivalently, if we
compactify on $Y$, the low-energy effective action will have $N=2$
supersymmetry and admits a flat supersymmetric ground state
$\bbR^{1,3}$. 

The condition~\eqref{ccons} really splits into two parts: first the
existence of a non-vanishing globally defined spinor $\eta$ on $Y$ and
second that $\eta$ is covariantly constant. The first condition
implies the existence of two four-dimensional supersymmetry parameters
and hence that the effective action has $N=2$ supersymmetry. The
second condition that $\eta$ is covariantly constant implies that the
effective action has a flat supersymmetric ground state.

The existence of $\eta$ is equivalent to the statement that the
structure group of the tangent bundle is reduced. To see what this
means, recall that the structure group  refers to the group of
transformations required to patch the tangent bundle (or more
precisely the bundle of orthonormal frames) over the manifold.  Thus
on a spacetime of the form $\bbR^{1,3}\times Y$ the structure group
reduces from $\SO(1,9)$ to $\SO(1,3)\times\SO(6)$ and the spinor
representation decomposes accordingly as
$\rep{16}\to\rep{(2,4)}+\rep{(\bar 2,\bar 4)}$. Suppose now that the
structure group of $Y$ reduces further to
$\SU(3)\subset\SO(6)\cong\SU(4)$. The $\rep{4}$ then decomposes as
$\rep{3}+\rep{1}$ under the $\SU(3)$ subgroup. An invariant spinor
$\eta$ in the singlet representation of $\SU(3)$ thus depends
trivially on the tangent bundle of $Y$ and so is globally defined and
non-vanishing. Conversely, the existence of such a globally defined
spinor implies that the structure group of $Y$ is $\SU(3)$ (or a
subgroup thereof). Mathematically, one says that the $Y$ has
$\SU(3)$-structure. In appendix~\ref{acs} we review some of the
properties of such manifolds from a more mathematical point of view
and for a more detailed discussion we refer the reader to the
mathematics literature~\cite{FFS,salamonb,joyce,friedrich,salamon,CS}. 
Here we will
concentrate on the physical implications.

The second condition that $\eta$ is covariantly constant has well
known consequences (as reviewed for instance in~\cite{GSW}). It is
equivalent to the statements that the Levi--Civita connection has
$\SU(3)$ holonomy or similarly that $Y$ is Calabi--Yau. It implies that an
integrable complex structure exists and that the corresponding
fundamental two-form $J$ is closed. In addition, there is a unique
closed holomorphic three-form $\Omega$. Together these structure and
integrability conditions imply that Calabi--Yau manifolds are complex,
Ricci-flat and K\"ahler.

Symmetry with the low-energy IIB theory with $H_3$-flux, implies that
compactification on generalized mirror manifold $\hat Y$ still leads to an
effective action that is $N=2$ supersymmetric. However, the IIB theory
with flux in general no longer has a flat-space ground state which
preserves all supercharges~\cite{TV,PM1,CKLT}. From the above discussion, we
see that this implies that we still have a globally defined
non-vanishing spinor $\eta$, but we no longer require that $\eta$ is
covariantly constant, so $\nabla_m \eta\neq0$. 
In other words, $\hat{Y}$ has $\SU(3)$-structure but generically the
Levi--Civita connection no longer has $\SU(3)$-holonomy, so in
general, $\hat{Y}$ is not Calabi--Yau. In particular, as discussed in
appendix~\ref{Rhf}, generic manifolds with $\SU(3)$-structure are not
Ricci flat. 

In analogy with Calabi--Yau manifolds
let us first use the existence of the globally defined
spinor $\eta$  to define other invariant
tensor fields.\footnote{For Calabi--Yau manifolds these constructions are
  reviewed, for example, in ref.~\cite{GSW}. For compactifications
  with torsion  they are generalized in ref. \cite{dWS,AS1,Hull1,KMPT} and
  here we closely follow these references.}
Specifically, one has a fundamental two-form
\begin{equation}
\label{defJ}
    J_{mn} = -i \eta^\dagger \Gamma_7 \Gx_{mn} \eta\ ,
\end{equation}
and a three-form
\begin{equation}\label{defO}
   \Omega =  \Omega^+ + i \Omega^-\ ,
\end{equation}
where
\begin{equation}
   \Omega^+_{mnp} = -i \eta^\dagger \Gx_{mnp} \eta \ ,\qquad
   \Omega^- _{mnp} = - \eta^\dagger \Gx_7\Gx_{mnp} \eta \ .
\end{equation}
By applying Fierz identities one shows
\begin{equation}
\label{eq:JOmegaIds}
\begin{gathered}
   J\wedge J \wedge J
      = \frac{3\ii}{4}\, \Omega \wedge \bar{\Omega} \ , \\
   J \wedge \Omega = 0 \ ,
\end{gathered}
\end{equation}
exactly as for Calabi--Yau manifolds.
Similarly, raising an index on $J_{mn}$ and assuming a 
normalization $\eta^\dag\eta=1$, one finds 
\begin{equation}\label{Jac}
J_m{}^p J_p{}^n = - \dx_m{}^n\ , \qquad
J_m{}^p J_n{}^r g_{pr} = g_{mn}\ ,
\end{equation}
by virtue of the $\Gamma$-matrix algebra. This implies that $J_m{}^p$
defines an almost complex structure such that the metric $g_{mn}$ is
Hermitian with respect to $J_m{}^p$. 
The existence of an almost complex structure is sufficient 
to define $(p,q)$-forms as we review in  appendix~\ref{acs}. In
particular, one can see that $\Omega$ is an $\SU(3)$-invariant
$(3,0)$-form.  

Thus far we have used the existence of the $SU(3)$-invariant spinor $\eta$ to
construct $J$ and $\Omega$. One can equivalently characterize manifolds
with $SU(3)$-structure by the existence of a globally defined,
non-degenerate two-form $J$ and a globally defined non-vanishing complex
three-form $\Omega$ satisfying the conditions~\eqref{eq:JOmegaIds}.
Together these then define a metric \cite{joyce,H}.

The key difference from the Calabi--Yau case is that 
a generic $\hat Y$ does not have $SU(3)$ holonomy
since $\nabla_m\eta\neq0$. Using  \eqref{defJ}  and \eqref{defO}
this immediately implies that
$J$ and $\Omega$ are also generically no longer covariantly constant
$\nabla_m J_{np} \neq 0 \ ,  \nabla_m \Omega_{npq} \neq 0$. 
The deviation from being covariantly constant
is a measure of the deviation from $SU(3)$ holonomy
and thus a measure of the deviation from the Calabi--Yau condition.
This can be made more explicit by using the fact that 
on $\hat Y$ there always exists another connection 
$\nabla^{(T)}$, which is metric compatible
(implying $\nabla_m^{(T)}g_{np}=0$), and which does
satisfy $\nabla_m^{(T)} \eta = 0$
\cite{salamonb,joyce}.
The difference between any two metric-compatible
connections is a tensor, known as the contorsion $\CT_{mnp}$,
and thus we have explicitly
\begin{equation}
\label{torKS}
\nabla_m^{(T)} \eta = \nabla_m \eta-\frac14\CT_{mnp}\Gx^{np} \eta = 0\ ,
\end{equation}
where $\Gx^{np}$ is the antisymmetrized product of $\Gamma$-matrices
defined in appendix~\ref{conv} and $\CT_{mnp}$ takes
values in $\Lambda^1\otimes\Lambda^2$ ($\Lambda^p$ being  the space of
$p$-forms). We see that $\CT_{mnp}$ is the obstruction to $\eta$
being covariantly constant with respect to the Levi-Civita connection
and thus for non-vanishing $\CT$ the manifold $\hat Y$ can not
be Calabi--Yau. Similarly, using \eqref{defJ}, \eqref{defO}
and \eqref{torKS} one shows that
$J$ and $\Omega$ are also generically no longer covariantly constant but
 instead obey
\begin{equation}
\label{torJO}
\begin{aligned}
   \nabla_m^{(T)} J_{np} &= 
      \nabla_m J_{np} - \CT_{mn}{}^r J_{rp} - \CT_{mp}{}^r J_{nr} 
      = 0 \ , \\
   \nabla_m^{(T)} \Omega_{nmp} &= 
      \nabla_m \Omega_{npq} - \CT_{mn}{}^r\Omega_{rpq} 
         - \CT_{mp}{}^r\Omega_{nrq} - \CT_{mq}{}^r\Omega_{npr}
      = 0 \ , 
\end{aligned}
\end{equation}
so again $\CT$ is measuring the obstruction to $J$ and $\Omega$
being covariantly constant with respect to the Levi-Civita connection. 
We see that the connection $\nabla^{(T)}$ preserves the $SU(3)$ structure
in that $\eta$ or equivalently $J$ and $\Omega$ are constant with
respect to $\nabla^{(T)}$.

Let us now analyze the contorsion $\CT\in\Lambda^1\otimes\Lambda^2$ in
a little more detail.  Recall that $\Lambda^2$ is isomorphic to the
Lie algebra $\so(6)$, which in turn decomposes into $\su(3)$ and
$\su(3)^\perp$, with the latter defined by
$\su(3)\oplus\su(3)^\perp\cong\so(6)$. Thus the contorsion actually
decomposes as $\CT^{\su(3)}+\CT^0$ where
$\CT^{\su(3)}\in\Lambda^1\otimes\su(3)$ and
$\CT^0\in\Lambda^1\otimes\su(3)^\perp$. Consider now the action of
$\CT$ on the spinor $\eta$. Since $\eta$ is an $\SU(3)$ singlet, the
action of $\su(3)$ on $\eta$ vanishes, and thus, from~\eqref{torKS},
we see that  
\begin{equation}
\label{eq:CTdef}
   \nabla_m \eta = \frac14 \CT^0_{mnp} \Gamma^{np}\eta\ .
\end{equation}
{}From~\eqref{torJO}, one finds that analogous expressions hold for
$\nabla_mJ_{np}$ and $\nabla_m\Omega_{npq}$. We see that the
obstruction to having a covariantly constant spinor (or equivalently
$J$ and $\Omega$) is actually measured by not the full contorsion
$\CT$ but by the so-called ``intrinsic contorsion'' part $\CT^0$. 
Eq.~\eqref{eq:CTdef} implies that $\CT^0$ is independent
of the choice of $\nabla^{(T)}$ satisfying~\eqref{torKS}, and thus
is a property only of the $\SU(3)$-structure. This fact is reviewed in more
detail in appendix~\ref{acs}.

Mathematically, 
it is sometimes more conventional to use the torsion $T$ instead
of the contorsion $\CT$; the two are related via
$T_{mnp}=\frac12(\CT_{mnp}-\CT_{nmp})$ and  $T_{mnp}$ 
also satisfies \eqref{RT}. 
Similarly, one usually refers to the corresponding ``intrinsic torsion''
$T^0_{mnp}=\frac12(\CT^0_{mnp}-\CT^0_{nmp})$ which also is an element
of $\Lambda^1\otimes\su(3)^\perp$ and is in one-to-one correspondence
with $\CT^0$.\footnote{%
Note that our terminology will not be very precise in that often we will use
the term torsion when in fact we mean by this intrinsic torsion.}
If $\CT^0$ and hence $T^0$ vanishes, we say that the
$\SU(3)$ structure is torsion-free. This implies $\nabla_m\eta=0$ and
the manifold is Calabi--Yau.

Both $\CT^0$ and $T^0$  can be decomposed
in terms of irreducible $SU(3)$ representations and hence
different $\SU(3)$ structures 
can be characterized by the non-trivial $SU(3)$ representations
$T^0$ carries. Adopting the
notation used in~\cite{salamon,CS} we denote this decomposition by
\begin{equation} 
  \label{TinW}
  T^0 \in \W_1 \oplus \W_2  \oplus \W_3 \oplus \W_4 \oplus \W_5\ ,
\end{equation}
with the corresponding parts of $T^0$ labeled by $T_i$ with
$i=1,\dots,5$ and where the representations corresponding to the
different $\W_i$ are given in table \ref{tabW}. 

\begin{table}[htbp]
  \begin{center}
    \begin{tabular}[h]{|c|c|c|}
      \hline
      component & interpretation & $SU(3)$-representation \\
      \hline
      $\W_1$ & $J\wg \dd\Ox$\ \ or\ \ $\Omega\wg\dd J$ &
           $\rep 1 \oplus \rep{1}$ \\
      \hline
      $\W_2$ & $(\dd \Ox)^{2,2}_0$ & $\rep 8 \oplus \rep{8}$ \\
      \hline
      $\W_3$ & $ (\dd J)^{2,1}_0+(\dd J)^{1,2}_0$ &
         $\rep 6 \oplus \rep{\bar 6}$ \\
      \hline
      $\W_4$ & $J \wg \dd J$ & $\rep 3 \oplus \rep{\bar 3}$ \\
      \hline
      $\W_5$ & $\dd \Ox^{3,1}$
         & $\rep 3 \oplus \rep{\bar 3}$ \\
      \hline
    \end{tabular}
    \caption{The five classes of the intrinsic torsion of a space with
      $\SU(3)$ structure.}
    \label{tabW}
  \end{center}
\end{table}

The second column of table~\ref{tabW}, gives an interpretation of each
component of $T^0$ in terms of exterior derivatives of $J$ and
$\Omega$. The superscripts refer to projecting onto a particular
$(p,q)$-type, while the $0$ subscript  refers to the irreducible
$\SU(3)$ representation with any trace part proportional to $J^n$
removed (see appendix~\ref{app:IT}). This interpretation arises since,
from~\eqref{torJO}, we have 
\begin{equation}
\label{dJO}
\begin{aligned}
   \dd J_{mnp} &= 6 T^0_{[mn}{}^r J_{r|p]} \ , \\
   \dd \Ox_{mnpq} &= 12 T^0_{[mn}{}^r \Ox_{r|pq]}\ .   
\end{aligned}
\end{equation}
These can then be inverted to give an expression for each component
$T_i$ of $T^0$ in terms of $\dd J$ and $\dd\Omega$. This is discussed
in more detail from the point of view of $SU(3)$ representations in
appendix~\ref{app:IT}.  

Manifolds with $SU(3)$ structure are in general not complex
manifolds. An almost complex structure $J$ (obeying \eqref{Jac})
necessarily exists but the integrability of $J$ is 
determined by the vanishing of the Nijenhuis tensor $N_{mn}{}^p$.
{}From its definition~(\ref{Ntens}) we see that a covariantly constant
$J$ has a vanishing $N_{mn}{}^p$ and in this situation the manifold
is complex and K\"ahler (as is the case for Calabi--Yau manifolds). 
However, for a generic $J$  the Nijenhuis tensor does not vanish
and is instead determined by the (con-) torsion using 
(\ref{Ntens}) and (\ref{torJO}). Thus $T^0$ also is an obstruction to
$\hat Y$ being a complex manifold. However, one can show~\cite{salamon,CS} 
that $N_{mn}{}^p$ does not depend on all torsion components but is
determined entirely by the component of the torsion
$T_{1\oplus2}\in\W_1\oplus\W_2$, through  
\begin{equation}
\label{NIT}
   N_{mn}{}^p = 8 (T_{1\oplus2})_{mn}{}^p \ . 
\end{equation}

Before we proceed let us summarize the story so far. The requirement
of an $N=2$ supersymmetric effective action led us to consider
manifolds $\hat{Y}$ with $SU(3)$-structure. Such manifolds admit a
globally defined $SU(3)$-invariant spinor $\eta$ but the holonomy group of
the Levi-Civita connection is no longer $SU(3)$. The deviation from
$SU(3)$ holonomy is measured by the intrinsic (con-)torsion, and
implies that generically the manifold is neither complex nor
K\"ahler. However, the fundamental two-form $J$ and the $(3,0)$-form
$\Omega$ can still be defined; in fact their existence is equivalent
to the requirement that $\hat{Y}$ has $\SU(3)$-structure. 
Different classes of manifolds with $SU(3)$ structure are
labeled by the $SU(3)$-representations in which the intrinsic torsion
tensor resides. In terms of $J$ and $\Omega$ this is measured by which
components of the exterior derivatives $\dd J$ and $\dd\Omega$ are
non-vanishing.

\subsection{Half-flat manifolds}
\label{sec:hf}

In general, we might expect that there are further restrictions on
$\hat{Y}$ beyond the supersymmetry condition that it has
$\SU(3)$-structure. This would correspond to constraining the
intrinsic torsion so that only certain components in table
\ref{tabW} are non-vanishing. We provide evidence for  a particular
set of constraints in the following subsections.
Then, in section~\ref{KK}, we verify that these conditions do lead to
the required mirror symmetric type IIA effective action.

Before doing so, however, let us consider two arguments suggesting how
these constraints might appear.  First, recall that the K\"{a}hler
moduli on the Calabi--Yau manifold are paired with the $B_2$ moduli as
an element $B_2+\ii J$ of $H^2(Y,\bbC)$ where $J$ is the K\"{a}hler
form. Under mirror symmetry, these moduli map to the complex structure
moduli of $\tilde{Y}$ which are encoded in the closed holomorphic
$(3,0)$-form $\Omega$. 
Turning on $H_3$ flux on the original Calabi--Yau manifold means that
the real part of the complex K\"{a}hler form $B_2+\ii J$ is no longer
closed. Under the mirror symmetry, this suggests that we now have a
manifold $\hat{Y}$ where half of $\Omega=\Omega^++\ii\Omega^-$, in
particular $\Omega^+$, is no longer closed. 
From table~\ref{tabW}, we see that $\dd\Omega^{2,2}$ is related to 
the classes $\W_1$ and $\W_2$ which can be further decomposed into
$\W^+_1\oplus\W^-_1$ and $\W^+_2\oplus\W^-_2$ giving 
\begin{equation}
\label{T+-}
\begin{aligned}
   T^+_{1\oplus2} &\text{ corresponding to }
      (\dd\Omega^+)^{2,2} \ , \\
   T^-_{1\oplus2} &\text{ corresponding to }
      (\dd\Omega^-)^{2,2} \ .
\end{aligned}
\end{equation}
Thus, the above result that only $\Ox^-$ remains closed suggests that, 
\begin{equation}
\label{eq:T-}
   T_{1\oplus 2}^- = 0 \ .
\end{equation}
One might expect that it also implies that half of the $\W_5$
component vanishes. However, as discussed in~\cite{CS},
$(\dd\Omega^+)^{3,1}$ and $(\dd\Omega^-)^{3,1}$ are related, so, in
fact, all of the component in $\W_5$ vanishes and we have in addition
\begin{equation}
   T_5 = 0 \ .
\end{equation}

The second argument comes from the fact that the intrinsic torsion $T^0$
should be such that it supplies the missing $2(h^{(1,1)}+1)$ NS-fluxes.
In other words we need the new fluxes to be counted by the even
cohomology of the original Calabi--Yau manifold $Y$. This implies that
there should be some well-defined relation between $\hat Y$ and the
Calabi--Yau manifold $Y$. We return to this relation in more detail in
section~\ref{spectrum} but here let us simply make the very naive
assumption that we try to match the $\SU(3)$ representations of the
$H^{p,q}(Y)$ cohomology group with the $\SU(3)$ representations of
$T^0$. This suggests setting
\begin{equation}
   T_4 = T_5 = 0 \ .
\end{equation}
since the corresponding $H^{3,2}(Y)$ and $H^{3,1}(Y)$ groups vanish on
$Y$. On the other hand $T_{1,2,3}$ can be non-zero as the
corresponding cohomologies do exist on $Y$.

Taken together, these arguments suggest that the appropriate
conditions might be
\begin{equation}
   T_{1\oplus2}^- = T_4 = T_5 = 0 \ .
\end{equation}
This is in fact a known class of manifolds, denoted
\textit{half-flat}~\cite{CS}. From table~\ref{tabW} it is easy to see
that the necessary and sufficient conditions can be written as
\begin{equation}
\label{eq:hf}
\begin{aligned}
   \dd \Omega^- &= 0 , \\
   \dd \left( J \wedge J \right) &= 0 .
\end{aligned}
\end{equation}

It will be useful in the following to have explicit expressions for
the components of the intrinsic torsion $T_1$, $T_2$ and $T_3$ which
are non-vanishing when the manifold is half-flat. From table
\ref{tabW} we recall that $T_{1\oplus2}$ is in the same $\SU(3)$
representation as a complex four-form $F^{(2,2)}$ of type
$(2,2)$. Explicitly we have
\begin{equation}
  \label{cxT12}
  (T_{1\oplus 2})_{mn}{}^p = 
      F_{mnrs} \Ox^{rsp} + \bar F_{mnrs} \bar \Ox^{rsp} \ .
\end{equation}
The half flatness condition $T_{1\oplus 2}^- = 0$ just imposes that
$F$ is real ($F=\bar F$) so that
\begin{equation}
\label{su3Nt}
   (T_{1\oplus2})_{mn}{}^p\ = \ (T_{1\oplus2}^+)_{mn}{}^p\ =  \
   2F_{mnrs}^{(2,2)}\, \Ox^{+rsp} \ , 
\end{equation}
where we have used \eqref{defO}. Explicitly, from the
relations~\eqref{dJO} one has that $F$ is related to $\dd\Omega$
by\footnote{%
  Note, that up to this point, the normalization~$\eta^\dag\eta=1$
  fixed the normalization of $J$ and $\Omega$. In the following it
  will be useful to allow an arbitrary normalization of $\Omega$, thus
  we have included in this expression the general factor
  $||\Ox||^2\equiv\frac{1}{3!}\Ox_{\ax\bx\cx}\bar\Ox^{\ax\bx\cx}$. }
\begin{equation}\label{F22}
 F_{mnrs}^{(2,2)} 
     \equiv \frac{1}{4||\Omega||^2}\, (\dd\Omega)^{2,2}_{mnrs}
     = \frac{1}{4||\Omega||^2}\, (\dd\Omega^+)^{2,2}_{mnrs}\ .
\end{equation}
We will see in section~\ref{KK} that this plays the role of the NS
four-form which precisely complexifies the RR 4-form background flux
in the low-energy effective action.  This fact was anticipated
in~\cite{Vafa}. However, it will only generate the electric fluxes
defined in \eqref{H3flux},
i.e.~half of the missing NS-fluxes. As we said in the introduction,
the treatment of the magnetic fluxes, corresponding to the NS two-form
flux is more involved and will be discussed in a separate publication
\cite{GLMW}. 

Similarly, we see from table~\ref{tabW} that the component $T_{3}$ of
the torsion is in the same representation as a real traceless
three-form $A^{(2,1)}_0+\bar{A}^{(1,2)}_0$ of type $(2,1)+(1,2)$ (see
also appendix~\ref{app:IT}). From~\eqref{dJO} we see that this form is
proportional to $(\dd J)^{(2,0)}_0$. Explicitly we have 
\begin{eqnarray}
  \label{T3}
  (T_3)_{mnp} = \frac14 \Big(\dx_m^{m'} \dx_n^{n'} - J_m{}^{m'}
J_n{}^{n'}
  \Big) J_p{}^{p'} (\dd J)_{m'n'p'} - 2 F\, (\Ox^+)_{mnp} \ ,
\end{eqnarray}
where by $F$ we denoted the trace of $F_{mnpq}$ defined in \eqref{TrF}.

The remainder of the section focuses on providing evidence that equations
\eqref{eq:hf} are indeed the correct conditions.
Before doing so, recall that compactifications on manifolds with
torsion have also been discussed in
refs.~\cite{AS1,Hull1,dWS,BD,friedrich,rocek,papa,waldram}. The 
philosophy of these papers was slightly different in that
they consider backgrounds where in addition some of the $p$-form field
strength were chosen non-zero and one still satisfied $\delta \psi_m =0$.
Here instead we want the torsion to generate terms which mimic or rather
are mirror symmetric to NS-flux backgrounds. As a consequence, one
finds rather different conditions. Since in both cases one wants $N=2$
supersymmetry in four dimensions, the class of manifolds discussed
in~\cite{AS1,Hull1,dWS,BD,friedrich,rocek,papa} are also manifolds with
$SU(3)$ structure. However, in these cases the torsion is a traceless
real three-form. This implies $T \in \W_3 \oplus \W_4 \oplus \W_5$, but with
the $T_4$ and $T_5$ being no longer independent.\footnote{
  After the completion of this paper we received \cite{CCDLMZ} which discusses
  this issue in more detail. We thank K. Dasgupta for pointing out an error in
  the earlier version of the paper.} 
Thus $T_1=T_2=0$ and as a consequence the Nijenhuis tensor vanishes (since
it depends only on $T_{1\oplus2}$) and the manifolds are complex but
not K\"ahler.


\subsection{The complex three-torus and the SYZ picture}
\label{sec:torus}

Obviously the most direct approach to finding the structure of
$\hat{Y}$ is to consider a Calabi--Yau manifold where we can do the
mirror symmetry explicitly. The simplest example is $T^6$ viewed as a
complex three-torus where mirror symmetry is realized by T-duality on
$T^3\subset T^6$. This case we study explicitly in this section and our
discussion overlaps with refs.\ \cite{VW,dh,kachru} where also
orbifolds of $T^6$ are considered.
Furthermore, as we note at the end of the
section, given the SYZ conjecture~\cite{SYZ}, which argues that mirror
manifolds can be realized as $T^3$ fibrations, one also gets a picture of
how the analysis generalizes to arbitrary Calabi--Yau manifolds. 

For a square complex three-torus with unit length sides we can write
the metric as 
\begin{equation}
\label{eq:T6}
   \dd s^2 = \dd z^1\dd\bar{z}^1 + \dd z^2\dd\bar{z}^2
      + \dd z^3\dd\bar{z}^3 \ ,
\end{equation}
where we have chosen a complex structure $\dd z^\alpha=\dd x^\alpha+\ii\dd
y^\alpha$ for $\alpha=1,2,3$. The K\"ahler form and holomorphic
three-form are given by 
\begin{equation}
\begin{aligned}
   J &= - \frac{\ii}{2} \; \delta_{\alpha\bar{\beta}}
      \dd z^\alpha \wedge \dd\bar{z}^{\bar{\beta}} \ , \\
   \Omega &= \dd z^1 \wedge \dd z^2 \wedge \dd z^3 \ .
\end{aligned}
\end{equation}
Mirror symmetry then corresponds to doing three T-dualities in the
$x^\alpha$ directions.

We want to start with some $H_3\in H^3(T^6,\bbR)$ flux on the
torus. Because the torus is such a trivial example of a Calabi--Yau
manifold, its cohomology does not have the standard form. For
example, on a true Calabi--Yau threefold $H^1(Y,\bbR)=0$. However this
is clearly not the case on $T^6$. More relevant to us is that for
a Calabi--Yau manifold any element $H_3\in H^3(T^6,\bbR)$ is
primitive, meaning that $J\wedge\gamma=0$. Thus to match the generic
case we should ensure that the flux $H^3$ is primitive. (An equivalent
statement is that the primitive elements are the ones which survive
orbifolding the torus to give a true Calabi--Yau manifold.)

The second point to bear in mind is that T-duality is only a
symmetry of consistent string backgrounds, or, more simply, solutions of
the IIA and IIB supergravity equations. However the space
$\bbR^{1,3}\times T^6$ with non-zero $H_3=\gamma\in H^3(T^6,\bbR)$ is
not such a solution. Nonetheless it is easy to construct suitable
solutions by viewing the flux as coming from a wrapped NS
five-brane. As we will see, finding the T-dual solution is then simply
a smeared version of the general result~\cite{ov} that the transverse
T-dual of $k+1$ NS five-branes is an ALE space with an $A_k$
singularity.

Recall that the five-brane solution in flat space is given by~\cite{NSfive}
\begin{equation}
\label{eq:NS5}
\begin{aligned}
   \dd s^2 &= \dd s^2_{\bbR^{1,5}} + V \dd s^2_{\bbR^4} \ , \\
   H_3 &= *_4 \dd V\ , \\
   \ee^{2\Phi} &= V\ ,
\end{aligned}
\end{equation}
where $\dd s^2_{\bbR^{1,5}}$ describes the flat worldvolume, $V\dd
s^2_{\bbR^4}$ the conformally flat space transverse to the brane and
$*_4$ is the Hodge star on $\dd s^2_{\bbR^4}$. The function $V$ is
harmonic in the same transverse four-dimensional flat space. Smearing
the five-brane in three of the transverse directions corresponds to
solution where the harmonic function depends on only one
coordinate. We write $V=\lambda\xi$ with $\lambda$ constant and $\dd
s^2_{\bbR^4}=\dd\xi^2+(\dd\eta^1)^2+(\dd\eta^2)^2+(\dd\eta^3)^2$. Let
us similarly split off three of the worldvolume directions so $\dd
s^2_{\bbR^{1,5}}=\dd
s^2_{\bbR^{1,2}}+(\dd\eta^4)^2+(\dd\eta^5)^2+(\dd\eta^6)^2$. The
solution can then be written as 
\begin{eqnarray}
\label{eq:dmT6}
   \dd s^2 &=& \dd s^2_{\bbR^{1,2}} + V \dd\xi^2
      + \left( V(\dd\eta^1)^2 + V(\dd\eta^2)^2+ V(\dd\eta^3)^2
         + (\dd\eta^4)^2 + (\dd\eta^5)^2+ (\dd\eta^6)^2 \right) \ ,\nonumber\\
   H_3 &=& \lambda \dd\eta^1 \wedge \dd\eta^2 \wedge \dd\eta^3\ , \\
   \ee^{2\Phi} &=& V = \lambda\xi \ .\nonumber
\end{eqnarray}
Since this is invariant under translations of all the $\eta^i$
coordinates, these directions can be compactified to form a six-torus. 
We see that the three-form flux is entirely on this internal
$T^6$. In the non-compact four-dimensional space we have a (singular)
domain wall located at $\xi=0$ with a linear dependence on $\xi$. (To
include the source one takes $V=\lambda|\xi-\xi_0|$, giving the domain
wall at $\xi=\xi_0$.) Geometrically the solution is a $T^6$ fibration
over the half line $\bbR^+$ parameterized by $\xi$. Physically, we
have a ``stack'' of five-branes all wrapping the
$(\eta^4,\eta^5,\eta^6)$ torus and smeared in the
$(\eta^1,\eta^2,\eta^3)$ directions on the $T^6$. Since two spatial
directions of the five-branes are unwrapped, in the non-compact
four-dimensional space they appear as domain walls. 

As we discuss further below, these solutions generalize to the case of
$H_3$ flux on an arbitrary Calabi--Yau manifold $Y$, appearing as BPS
domain wall solutions of the four-dimensional effective
action~\cite{CKLT,BGS,BCL}.

We are now in a position to consider the action of mirror symmetry on
a solution of the form~\eqref{eq:dmT6}. There are several ways we
could identify the complex structure in~\eqref{eq:dmT6}. As an example
let us take $(x^1,x^2,x^3)=(\eta^4,\eta^5,\eta^1)$ and
$(y^1,y^2,y^3)=(\eta^2,\eta^3,\eta^6)$. Thus the flux is given by the
primitive form
\begin{equation}
   H_3 = \lambda \dd y^1 \wedge \dd y^2 \wedge \dd x^3\ .
\end{equation}
Mirror symmetry is the same as T-duality in the $x^\alpha$
directions. We can realize this explicitly by first choosing a gauge
where locally $B_2=\lambda y^1\dd y^2\wedge\dd x^3$ independent of the
$x^\alpha$ coordinates. T-duality in the $x^1$ and $x^2$ directions is
then essentially trivial, simply inverting the size of the $x^1$ and
$x^2$ circles. From the usual formulae derived in ref.~\cite{buscher}, 
the mirror solution is given by 
\begin{equation}
\begin{split}
\label{eq:met}
   \dd s^2 &= \dd s^2_{\bbR^{1,2}} + V \dd z^2 \\
      & \qquad + \left[ (\dd x^1)^2 + (\dd x^2)^2
         + V^{-1} \left(\dd x^3 - \lambda y^1 \dd y^2\right)^2
         + V (\dd y^1)^2 + V (\dd y^2)^2+ (\dd y^3)^2 \right] ,
\end{split}
\end{equation}
with $\ee^{2\Phi}=1$ and $H_3=0$. (Note, this same
calculation of essentially the T-dual of flat space with constant
$H_3$-flux has been considered several times before. Recent related
examples are found in~\cite{dh,kachru}.)   

We see that we again have a domain wall solution, but now it is
completely geometrical, with no $H$-flux and a constant dilaton. It
also has the form of a fibration of a six-dimensional manifold
$\hat{Y}$ over $\bbR^+$. 
However, $\hat{Y}$ is not a torus. 
This matches our expectation: the mirror of $T^6$ with $H_3$ flux is
no longer a Calabi--Yau. We now turn to investigating what
structure $\hat{Y}$ has at any given fixed value of $V$.

Geometrically, $\hat{Y}$ has the form $T^3\times Q$ where $Q$ is a
$S^1$ fibration over $T^2$, with $x^3$ the coordinate on $S^1$ and
$y^1$ and $y^2$ the coordinates on $T^2$~\cite{chris}. (More generally
one can view it as a special case of a $T^3$ fibration over $T^3$.)
This immediately implies a quantization condition. For the fibration
to be properly defined, it is easy to see that
\begin{equation}
\label{eq:quant}
   \lambda \in \bbZ \ .
\end{equation}
Viewed as a $U(1)$ bundle over $T^2$, this is simply the statement
that the first Chern class must be integral. This is interesting,
since, it reproduces the quantization of the original flux $H_3\in
H^3(Y,\bbZ)$ as a string theory background.

Next we note that we can still introduce a candidate complex structure
on $\hat{Y}$. We have a basis of globally defined orthonormal
complex one-forms given by
\begin{equation}
\label{eq:holo}
\begin{aligned}
   e^1 &= \dd x^1 + \ii \sqrt{V} \dd y^1\ , \\
   e^2 &= \dd x^2 + \ii \sqrt{V} \dd y^2\ , \\
   e^3 &= \frac{1}{\sqrt{V}}\left(\dd x^3 - \lambda y^1 \dd y^2\right)
      + \ii \dd y^3\ ,
\end{aligned}
\end{equation}
However, clearly, this cannot be integrated to give complex
coordinates $z^\alpha$. Thus this in fact only defines an almost complex
structure. We can define the associated K\"{a}hler form
\begin{equation}
   J = - \frac{\ii}{2} \;  \delta_{\alpha\bar{\beta}} 
      e^\alpha \wedge \bar{e}^{\bar{\beta}}\ .
\end{equation}
We immediately see that (taking the exterior derivative on $\hat{Y}$
only)
\begin{equation}
\label{eq:Jcond1}
   \dd J = -\frac{2\lambda}{\sqrt{V}}
       \dd y^1 \wedge \dd y^2 \wedge \dd y^3 \neq 0\ ,
\end{equation}
while on the other hand we do find
\begin{equation}
\label{eq:Jcond2}
   \dd(J\wedge J)=0\ .
\end{equation}
We can also globally define a holomorphic three-form
\begin{equation}
   \Omega = \Omega^+ + \ii \Omega^- = e^1 \wedge e^2 \wedge e^3 ,
\end{equation}
satisfying
\begin{equation}
   \dd \Omega = -\frac{\lambda}{\sqrt{V}}\;
      \dd x^1\wedge\dd x^2\wedge\dd y^1\wedge\dd y^2 ,
\end{equation}
so that
\begin{equation}
\label{eq:Ocond}
   \dd \Omega^+ \neq 0 , \qquad \dd \Omega^- = 0 .
\end{equation}

In summary, we see that, first, one can still define $J$ and $\Omega$
implying that $\hat{Y}$ does indeed have $\SU(3)$-structure as we
argued above was necessary for a low-energy $N=2$ effective
action. Secondly, this structure is not Calabi--Yau since $J$ and
$\Omega$ are not closed. Instead we have exactly the half-flat
conditions~\eqref{eq:hf} as suggested above. We note that the
T-duality analysis given here can be easily generalized to a class of
flux configurations on $T^6$ of the form
\begin{equation}
   H_3 = \lambda_1 \dd y^1 \wedge \dd y^2 \wedge \dd x^3
      +  \lambda_2 \dd y^2 \wedge \dd y^3 \wedge \dd x^1
      +  \lambda_3 \dd y^3 \wedge \dd y^1 \wedge \dd x^2 \ ,
\end{equation}
as well as more general tori, giving the same set of conditions. Thus,
we see that at least the corresponding sub-class of
generalized mirror manifolds $\hat{Y}$ are all half-flat.  

Finally, let us comment on how this picture might generalize to
arbitrary Calabi--Yau manifolds. Recall the SYZ picture of mirror
symmetry~\cite{SYZ}. This conjectures that any Calabi--Yau manifold
which has a mirror is a $T^3$ fibration with, in general,
singular fibers. Mirror symmetry is realized as T-duality on the
toroidal fibers. In particular, if we start in IIB with the manifold
$\tilde{Y}$, the moduli space of D3 branes wrapping the $T^3$ fiber of
$\tilde{Y}$ must be the same as the moduli space  of D0 branes on
the mirror IIA manifold $Y$. But this later space is just the manifold
$Y$ itself. Thus we can construct $Y$ from the moduli space of D3
branes wrapping the fibers. This space arises both from deformations
of the D3 in $\tilde{Y}$ and also the flat $U(1)$ connections on the
D3 brane. As such, classically, it is also a $T^3$ fibration over the
same base. (It also gets instanton corrections.) The complex torus
$T^6$ discussed here is a very simple example, realizing the SYZ
picture as a trivial $T^3$ fibration over $T^3$. Without flux, the
T-dual of $T^6$ is simply another six-torus. 

Now consider the case with flux. The point is that the NS two-form $B_2$
couples to the D3 brane in the Born--Infeld action. As a result the
moduli space is changed. Thus the mirror space is no longer $Y$, but
is a new manifold $\hat{Y}$. We saw this explicitly in the $T^6$
example. The generalized mirror manifold $\hat{Y}$ was again a $T^3$
fibration over $T^3$ but unlike $Y=T^6$ the fibration was no longer
trivial and hence the manifold was not Calabi--Yau. This suggests
that, in general, in the SYZ picture, the manifold $\hat{Y}$
corresponds to the original mirror manifold $Y$ with some ``extra
twists'' in the $T^3$ fibration, so that $\hat{Y}$ is not Calabi--Yau
or Ricci flat.  

Just as in~\cite{SYZ}, one can calculate the T-duality explicitly in
the large complex structure, or semi-flat, limit. In this limit the
$T^3$ fibers are very small compared with the size of the base
$\mathcal{B}$ of the fibration. Away from singular fibers, the metric
can then be written in a form which depends only on the coordinates
$y^i$ on $\mathcal{B}$
\begin{equation}
\label{semiflat}
   \dd s^2 = g_{ij}(y)\dd y^i\dd y^j
      + h^{\alpha\beta}(y)
         \left(\dd x_\alpha+\omega_\alpha(y)\right)
         \left(\dd x_\beta+\omega_\beta(y)\right) \ ,
\end{equation}
where $x_\alpha$ parameterize the $T^3$ fiber and $\omega_\alpha$ are
locally one-forms on $\mathcal{B}$ describing the twisting of the
fiber as one moves over the base. Metrics of this type are described
in~\cite{Hhf,gross}. They must satify certain conditions in order to be
Calabi--Yau. As in the $T^6$ example let us now consider a primitive
harmonic $H_3$-flux on the semi-flat metric of the form    
\begin{equation}
   H_3 = F_\alpha \wedge \dd x_\alpha \ ,
\end{equation}
where $F_\alpha$ are a triplet of harmonic two-forms on
$\mathcal{B}$. Locally, one can write $B_2=A_\alpha\wedge\dd x^\alpha$, where
$A_\alpha$ are the corresponding one-form potentials for
$F_\alpha$. In this gauge, the background is independent of $x^\alpha$
and one can make an explicit T-duality transformation along the
$T^3$. This generates a new metric of the same form
(it is actually related to it by a Legendre transform~\cite{Hhf})
except now with $\omega_\alpha$ replaced by
$\omega_\alpha+A_\alpha$. These new terms modify the twisting of
the $T^3$ fiber and mean that the metric is no longer
Calabi--Yau. This is precisely the ``extra twisting'' discussed above.

\subsection{Domain walls and fibered $G_2$ manifolds}
\label{sec:DM}

The above discussion can be generalized to arbitrary Calabi--Yau
compactifications in the following way. The point is that
$N=1$ supersymmetric domain wall solutions exist for any such
compactification with $H_3$ flux. 
This can be seen directly from the
low-energy effective action as discussed in~\cite{CKLT,BGS,BCL}. Just
as in the torus case, physically, one can view these solutions as NS
five-branes wrapped on special Lagrangian three-cycles on the
Calabi--Yau manifold. This leaves two unwrapped spatial dimensions and
hence corresponds to a BPS domain wall in four dimensions. As
ten-dimensional solutions, the five-branes are not localized in the
Calabi--Yau, and so the domain walls correspond in this sense to
five-branes smeared within the compact Calabi--Yau manifold.
By definition, compactifying IIA on $\hat Y$ leads to the same
effective action as compactifying IIB on $\tilde{Y}$ with flux
$H_3$. Thus the effective IIA theory on $\hat{Y}$ necessarily also
admits BPS domain wall solutions. As we will see, this requirement can
then be used to constrain the possible form of $\hat{Y}$.

From the point of view of the four-dimensional effective action the
$H_3$ flux provides a potential for essentially the complex structure
moduli describing $\Omega$, though, in fact, also for the dilaton
$\Phi$ and the K\"{a}hler modulus describing the overall size of the
Calabi--Yau manifold.
The domain walls then correspond to a solution where the moduli
depend non-trivially on the direction perpendicular to the wall. The
ten-dimensional solution in the string frame has the form
\begin{equation}
\label{eq:HDM}
\begin{aligned}
   \dd s^2 &= \dd s^2_{\bbR^{1,2}} + \dd y^2 + \dd s^2_Y(y) , \\
   \phi &= \phi(y) , \\  
   H_3 & \in H^3(Y,\bbR) ,
\end{aligned}
\end{equation}
where $y$ parameterizes the direction perpendicular to the wall, $\dd
s^2_{\bbR^{1,2}}$ is the flat metric on the worldvolume of the domain wall,
$\dd s^2_Y(y)$ is the metric on the Calabi--Yau $Y$, which through the
complex structure moduli and the overall volume is a function of $y$,
and the flux $H_3$ 
is a harmonic form in $H^3(Y,\bbR)$. Note this
is the same form as~\eqref{eq:dmT6} in the $T^6$ case above except we
made a change of variables from $z$ to $y$ for the transverse
coordinates to remove the factor multiplying $\dd z^2$. Geometrically,
the solution has the form  $\bbR^{1,2}\times Z$ where $Z$ is a
non-compact seven-dimensional manifold which is a fibration $Z\to I$
of the Calabi--Yau manifold $Y$ over an interval $I\subset\bbR$
parameterized by $y$. Again this is just as for the $T^6$ case.

Now consider the mirror of these domain wall solutions. The
four-dimensional effective actions will be the same, simply with the
role of the complex structure moduli and complexified K\"{a}hler
moduli exchanged. Thus there will still be supersymmetric domain wall
solutions breaking half the supersymmetries,  but now these arise from
a potential for the complexified K\"{a}hler moduli. 
Let us assume
that, as above, that the mirror compactification should be pure
geometrical with no $H_3$ flux and trivial dilaton.  
The mirror solution then still has the
domain wall form   
\begin{equation}
\label{eq:G2DM}
   \dd s^2 = \dd s^2_{\bbR^{1,2}} + \dd y^2 + \dd s^2_{\hat{Y}}(y) \ ,
\end{equation}
but now $H_3$ is zero and $\Phi$ is constant. Thus again we have the
structure $\bbR^{1,2}\times\hat{Z}$ where $\hat{Z}$ is a non-compact
seven-dimensional manifold which is a fibration $\hat{Z}\to I$ of
$\hat{Y}$ over an interval $I\subset\bbR$ parameterized by $y$.

Now we use the condition that the domain wall should break half the
supersymmetries. First recall that for the low-energy effective action
to be supersymmetric the manifold $\hat{Y}$ has to have $\SU(3)$
structure. This is equivalent to the existence of the forms $J$ and
$\Omega=\Omega^++\ii\Omega^-$ everywhere on $\hat{Y}$.
Then, for the domain wall solution~\eqref{eq:G2DM} to be BPS it must
describe a supersymmetric manifold. In particular, $\hat{Z}$ must 
have $G_2$-holonomy. (As is the case, in the $T^6$ example, for the
metric~\eqref{eq:met}.) There is now an obvious question: what are the
conditions on the six-dimensional manifolds $\hat{Y}$ with $\SU(3)$
structure for $\hat{Z}$ to have $G_2$-holonomy? 

This has been answered in a very interesting paper by Hitchin~\cite{H}
(see also~\cite{CS}).  The manifold $\hat{Z}$ is a $G_2$-manifold iff
$\hat{Y}$ is half-flat. Again we get the same conditions we
found in the $T^6$ example~\eqref{eq:hf}. (Note, that, $\Omega$ is
only defined up to an overall phase, thus whether the real or
imaginary part or some other combination is closed is purely a choice of
conventions. In~\cite{CS}, the real part is closed, while here we take the
imaginary part to match the conventions used in the $T^6$ example
above.)

This concludes our analysis of manifolds with $SU(3)$ structure and
in particular of half-flat manifolds. We identified them as 
promising candidates to supply the missing (electric) NS-fluxes which 
are demanded by mirror symmetry. In the next section we provide  
further evidence for this proposal by explicitly compactifying 
type IIA on half-flat manifolds $\hat Y$.


\section{The dimensional reduction on $\hat Y$}
\label{KK}

Before we launch into the details of the dimensional reduction, recall
that we are aiming at the derivation of a type IIA effective action
which is mirror symmetric to the type IIB effective action obtained
from compactifications on Calabi--Yau threefolds with (electric) NS
3-form flux $H_3$ turned on.  This effective theory is reviewed in
appendix~\ref{IIBNSflux} while the Calabi--Yau compactification of type
IIA without fluxes is recalled in appendix~\ref{IIAcomp}. As we have
stressed throughout, the central problem is that in IIA theory there
is no NS form-field which can reproduce the NS-fluxes which are the
mirrors of $H_3$ in the type IIB theory. Vafa suggested that the type
IIA mirror symmetric configuration is a different geometry where the
complex structure is no longer integrable~\cite{Vafa}, so that the
compactification manifold $\hat{Y}$ is not Calabi--Yau. In the
previous section we have already collected evidence that half-flat
manifolds are promising candidates for $\hat{Y}$. The additional flux
was characterized by the four-form $F^{(2,2)}\sim\dd\Omega^{2,2}$. The
purpose of this section is to calculate the effective action, in an
appropriate limit, for  type IIA compactified on a half-flat
$\hat{Y}$, and show that it is exactly equivalent to the known
effective theory for the mirror type IIB compactification with
electric flux.

The basic problem we face in this section is that so far we have no 
mathematical procedure for constructing a half-flat manifold
$\hat{Y}$ from a given
Calabi--Yau manifold $Y$. Instead we will give a set of rules 
for the structure of $\hat{Y}$ and the corresponding light spectrum
by using physical considerations and in particular using
mirror symmetry as a guiding principle. 
Specifically, we will write a set of two-, three- and
four-forms on $\hat{Y}$ which are in some sense ``almost
harmonic''. By expanding the IIA fields in these forms, we can then
derive the four-dimensional effective action which is equivalent to
the known mirror type IIB action.

\subsection{The light spectrum and the moduli space of $\hat{Y}$}

\label{spectrum}

To derive the effective four-dimensional theory we first have to
identify the light modes in the compactification such as the
metric moduli. Unlike the case of a conventional reduction on a
Calabi--Yau manifold, from the IIB calculation we know that the
low-energy theory has a potential~\eqref{potIIB} and so not all the
light fields are massless. In any dimensional reduction there is
always an infinite tower of massive Kaluza--Klein states, thus we need
some criterion for determining which modes we keep in the effective
action.   

Recall first how this worked in the type IIB case. One starts with a
background Calabi--Yau manifold $\tilde{Y}$ and makes a perturbative
expansion in the flux $H_3$. To linear order, $H_3$ only appears in
its own equation of motion, while it appears quadratically in the
other equations of motion, such as the Einstein and dilaton
equations, so, heuristically,
\begin{equation}
\label{eq:eom}
\begin{gathered}
   \nabla^m H_{mnp} = \dots \ , \\
   R_{mn} = H^2_{mn} + \dots \ .
\end{gathered}   
\end{equation}
In the perturbation expansion we first solve the linear
equation on $\tilde{Y}$ which implies that $H_3$ is harmonic. We then
consider the quadratic backreaction on the geometry of $\tilde{Y}$ and
the dilaton. The backreaction will be small provided $H_3$ is small
compared to the curvature of the compactification, set by the inverse size
of the Calabi--Yau manifold $1/\LCY$. Recall, however, that in
string theory the flux $\int_{\gamma_3}H$, where $\gamma_3$ is any
three-cycle in $\tilde{Y}$ is quantized in units of
$\alpha'$. Consequently $H_3\sim\alpha'/\LCY^3$ and so for a small
backreaction we require $H_3/\LCY^{-1}\sim\alpha'/\LCY^2$ to be small. In
other words, we must be in the large volume limit where the
Calabi--Yau manifold is much larger than the string length, which
anyway is the region where supergravity is applicable. The
Kaluza--Klein masses will be of order $1/\LCY$. The mass correction
due to $H_3$ is proportional to $\alpha'/\LCY^3$ and so is comparatively
small in the large volume limit. Thus in the dimensional
reduction it is consistent to keep only the zero-modes on $\tilde{Y}$
which get small masses of order $\alpha'/\LCY^3$ and to drop all the
higher Kaluza--Klein modes with masses of order $1/\LCY$.  

We would like to make the same kind of expansion in IIA and think of
the generalized mirror manifold $\hat{Y}$ as some small perturbation
of the original Calabi--Yau $Y$ mirror to $\tilde{Y}$ without
flux. The problem we will face throughout this section is that we do
not have, in general, an explicit construction of $\hat{Y}$ from
$Y$. Thus we can only give general arguments about the meaning of
such a limit. From the previous discussion we saw that it is the
intrinsic torsion $T^0$ which measures the deviation of $\hat{Y}$ from
a Calabi--Yau manifold. Thus we would like to think that in the limit
where $T^0$ is small $\hat{Y}$ approaches $Y$. The problem is, as we
saw for the simple complex torus example in section~\ref{sec:torus}, in
general $Y$ and $\hat{Y}$ have different topology. Thus, at the
best, we can only expect that $\hat{Y}$ approaches $Y$ locally in
the limit of small intrinsic torsion. Put another way, the torsion,
like $H_3$ is really ``quantized'' in the sense that, again as we saw
in the torus example, it is associated with topological twists in the 
SYZ fibration structure of $\hat{Y}$. Consequently, it cannot really
be put to zero, instead we can only try distorting the space to a
limit where locally $T^0$ is small and then locally the manifold looks
like $Y$. 

This can be made slightly more formal in the following way. It is a
general result~\cite{FFS} that the Riemann tensor of any manifold with
$\SU(n)$ structure has a decomposition as 
\begin{equation}
\label{Rdecomp}
   R = R_{\text{CY}} + R_\perp \ ,
\end{equation}
where the tensor $R_{\text{CY}}$ has the symmetry properties of the
curvature tensor of a true Calabi--Yau manifold, so that, for instance
the corresponding Ricci tensor vanishes. The orthogonal component
$R_\perp$ is completely determined in terms of $\nabla T^0$ and
$(T^0)^2$. (Note that the corresponding decomposition of the Ricci 
scalar in the half-flat case is calculated explicitly in
appendix~\ref{Rhf}.) From this perspective, we can think of $R_\perp$
as a correction to the Einstein equation on a Calabi--Yau manifold,
analogous to the $H_3^2$ correction in the IIB theory. In particular,
if $\hat{Y}$ is to be locally like $Y$ in the limit of small torsion,
we require 
\begin{equation}
   R_{\text{CY}}(\hat{Y}) = R(Y) \ .
\end{equation}

What, however, characterizes the limit where the intrinsic torsion is
small? Unlike the IIB case the string scale does not appear in
$T^0$. Typically both curvatures $R_{\text{CY}}$ and $R_\perp$ are of
order $1/\hat{L}^2$ where $\hat{L}$ is the size of $\hat{Y}$. Thus
making $\hat{Y}$ large will not help us. Instead, we must consider
some distortion of the manifold so that $R_\perp\ll
R_{\text{CY}}$. What this distortion might be is suggested by mirror
symmetry. We know that, without flux, a large radius $\tilde{Y}$ is
mapped to $Y$ with large complex structure. Thus we might expect that
we are interested in the large complex structure limit of
$\hat{Y}$. It is easy to see that this is what happens for the example
of the complex torus. In the half-flat metric~\eqref{eq:met} suppose
we now take the $x^\alpha$ torus to be of radius $L_x$ and the
$y^\alpha$ torus to be of radius $L_y$. The parameter $\lambda$ in
\eqref{eq:quant} is then quantized in units of $L_x/L_y^2$. The
intrinsic torsion, measured by $\dd J$ and $\dd\Omega$ is proportional
to $\lambda$ and so is suppressed by a factor of a power of $L_x/L_y$
compared with the mass scale set by the volume of $\hat{Y}$. In this
sense the intrinsic torsion is small when $L_x/L_y$ is small which is
precisely the large complex structure limit. 

In this limit, the conjecture is that $R_\perp(\hat{Y})$ becomes a
small perturbation, with a mass scale much smaller than the Kaluza--Klein
scale set by the average size of $\hat{Y}$. Thus, as in the IIB case,
at least locally, the original zero modes on $Y$ become approximate
massless modes on $\hat{Y}$ gaining a small mass due to the
non-trivial torsion. This suggests it is again consistent in this
limit to consider a dimensional reduction keeping only the
deformations of $\hat{Y}$ which correspond locally to zero modes of
$Y$. This holds both for the ten-dimensional gauge potentials given in
case without flux in~\eqref{fexpA} and the deformations of the metric
as in~\eqref{Jexp} and~\eqref{Oxz}. 

Having discussed the approximation, let us now turn to trying to
identify this light spectrum more precisely and characterizing how the
missing NS flux enters the problem. As discussed, it is the intrinsic
torsion of $\hat{Y}$ which characterizes the deviation of $\hat{Y}$
from a Calabi--Yau manifold therefore we expect that this encodes the
NS-flux parameters we are looking for. Mirror symmetry requires that
these new NS-fluxes are counted by the even cohomology of the ``limiting''
Calabi--Yau manifold $Y$. 
As we saw above, in the case of half-flat
manifolds this suggests that the real $(2,2)$-form $F\sim\dd\Omega$ on
$\hat{Y}$, introduced in~\eqref{su3Nt} and discussed by
Vafa~\cite{Vafa}, can be viewed as specifying some ``extra data"
on $Y$ which is  a harmonic form 
$\ff\in H^4(Y,\bbR)$ (or equivalently $H^2(Y,\bbR)$) measuring, at
least part of, the missing NS flux.

While we have no explicit construction of $\hat{Y}$ in terms of $Y$ and some
given flux $\ff$,
nonetheless, we expect, if mirror symmetry is
to hold, that for each pair $(Y,\ff)$ there is a unique half-flat
manifold $\hat{Y}_\ff$, so that there is a map
\begin{equation}
\label{eq:ident}
   (Y,\ff) \Leftrightarrow \hat{Y}_\ff \ ,
\end{equation}
where, in the limit of small torsion (large complex structure),
$Y$ and $\hat{Y}_\ff$ with the corresponding metrics
are locally the same. In fact, we can argue two more conditions. First,
the identification~\eqref{eq:ident} can be applied at each point in
the moduli space of $Y$ giving us, assuming uniqueness, a
corresponding moduli space of $\hat{Y}_\ff$. Furthermore, from the
torus example, we see that the type IIB $H_3$-flux only effected the
topology of $\hat{Y}$ in the sense that all points in the moduli space
of $\hat{Y}_\ff$ for given flux had the same topology. Thus we see
that, if mirror symmetry is to hold, the moduli space of metrics
$\mathcal{M}(Y)$ and $\mathcal{M}(\hat{Y}_\ff)$ of $Y$ and $\hat{Y}$
are locally the same    
\begin{equation}
   \mathcal{M}(\hat{Y}_\ff) = \mathcal{M}(Y) \ ,
      \qquad \text{for any given $\ff$} \ ,
\end{equation}
where $\ff$ only effects the topology of
$\hat{Y}$. This gives the full moduli space of all $\hat{Y}_\ff$ the
structure of an infinite number of copies of $\mathcal{M}(Y)$ labeled
by $\ff$.\footnote{We thank Ron Donagi for discussions on this point.}

More explicitly, the matching of moduli spaces means that for each
$(\Omega,J)$ on $Y$, since $\hat{Y}_\ff$ has $\SU(3)$ structure, we
have a unique corresponding $(\Omega,J)$ on $\hat{Y}$ and we must have
a corresponding expansion in terms of a basis of forms on $\hat{Y}$ 
\begin{equation}
\label{Oxz2}
\begin{aligned}
   \Ox &= z^A\, \ax_A - \cF_A\, \bx^A\ , 
     \qquad  A= 0,1,\ldots,h^{(1,2)}(Y)\ , \\
   J &= v^i\, \omega_i \ ,\qquad  i=1,\ldots, h^{(1,1)}(Y)\ ,
\end{aligned}
\end{equation}
where $z^A= (1,z^a)$ with $a=1,\ldots,h^{(1,2)}(Y)$ and the $z^a$ are
the scalar fields corresponding to the deformations of the complex
structure ($\cF_A$ is  defined in appendix \ref{IIAcomp}), while the
$v^i$  are the scalar fields corresponding to the K\"ahler
deformations. The key point here is that although $(\ax_A,\bx^A)$ form
a basis for $\Omega$ and the $\omega_i$ form a basis for $J$ they are
not, in general, harmonic, and thus are not bases for $H^3(\hat{Y})$ and
$H^{(1,1)}(\hat{Y})$. Locally, however, in the limit of small
intrinsic torsion, they should coincide with the harmonic basis
of $H^3(Y)$ and $H^{(1,1)}(Y)$ on $Y$. 
For $*J$ one has an analogous  expansion
in terms of four-forms on $\hat{Y}$ as in
\eqref{oxstar}
\begin{equation}
\label{*J}
   *J = 4\mathcal{K} g_{ij}\nu^i\tilde{\omega}^j \ ,
      \qquad  i=1,\ldots, h^{(1,1)}(Y)\ , 
\end{equation}
where, again, there is no condition on $\tilde{\omega}^i$ being
harmonic on $\hat{Y}$, but in the small torsion limit they again
locally approach harmonic forms on $Y$. 

The above expressions~\eqref{Oxz2} and~\eqref{*J} have been written in
terms of a prepotential $\mathcal{F}$ and a metric $g_{ij}$ 
which defines the metric on the moduli space just as for 
$Y$. If the
low-energy effective action is to be mirror symmetric we necessarily
have that the metrics on the moduli spaces
$\mathcal{M}(\hat{Y}_\ff)$ and $\mathcal{M}(Y)$ agree. This means
that the corresponding kinetic terms in the low-energy effective
action agree and implies the conditions 
\begin{equation}
\label{normYhat}
   \int_{\hat Y} \ox_i \wg \tilde \ox^j =  \dx_i^j \  ,  \quad
   \int_{\hat Y} \ax_A \wg \bx^B =  \dx_A^B\ ,
      \quad
  \int_{\hat Y} \ax_A \wg \ax_B = \int_{\hat Y} \bx^A \wg \bx^B = 0\ ,
\end{equation}
exactly as on $Y$ in (\ref{normH2}) and (\ref{norm}).

Now let us return to the flux and the restrictions implied
by $\hat{Y}_\ff$ being half-flat. Recall that we have argued that
the four-form $F^{(2,2)}\sim(\dd\Omega)^{2,2}$ corresponds to a
harmonic form $\ff\in H^4(Y,\bbZ)$ measuring the flux. Given the map
between harmonic four-forms on $Y$ and the basis $\tilde{\omega}^i$
introduced in~\eqref{*J}, we are naturally led to 
rewrite \eqref{F22} as 
\begin{equation}
\label{F4}
\begin{split}
  F_{mnpq}^{(2,2)} &\equiv 
        \frac{1}{4||\Ox||^2} (\dd\Omega)^{2,2}_{mnpq} \\
     &= \frac{1}{4||\Ox||^2}\, e_i\, \tox^i_{mnpq}\ ,
        \qquad i= 1, \ldots , h^{(1,1)}(Y) \ ,
\end{split}
\end{equation}
where the $e_i$ are
constants parameterizing the flux. Again, in the limit of small
torsion, locally $F$ is equivalent to a harmonic form on $Y$, namely
$\ff$.  

Inserting (\ref{Oxz2}) into (\ref{F4}), we have
\begin{equation}
   \dd\Omega = z^A \dd\alpha_A - \mathcal{F}_A \dd\beta^A = e_i\tox^i \ .
\end{equation}
However, we argued that the flux only effects the topology of
$\hat{Y}$ and does not depend on the point in moduli space. Thus, we
require that this condition is satisfied independent of the choice of
moduli $z^A=(1,z^a)$.
This is only possible if we have 
\begin{equation}
  \label{dax}
  \dd \ax_0 = e_i\, \tox^i\ , \qquad \dd \ax_a = \dd \bx^A = 0\ ,
\end{equation}
where $\ax_0$ is singled out since it is the only direction in
$\Omega$ which is independent of $z^a$.\footnote{%
Of course this corresponds to a specific choice
of the symplectic basis of $H^3$. It is the same choice which
is conventionally used in establishing the mirror map without fluxes.}   
Furthermore, inserting
(\ref{dax}) into (\ref{normYhat}) gives
\begin{equation}
  \label{dox1}
  e_i = \int \ox_i \wg d \ax_0 = - \int  \dd \ox_i \wg \ax_0 \ .
\end{equation}
Thus consistency requires
\begin{equation}
  \label{dox}
  \dd \ox_i = e_i \bx^0\  , \qquad \dd \tilde\omega^i = 0\ ,
\end{equation}
where the second equation follows from (\ref{dax}).\footnote{%
Strictly speaking also  $d \ox_i =
  e_i \bx^0 + a^A \ax_A + b_a \bx^a$ for some yet undetermined
  coefficients $a^A, b_a$ solves (\ref{dox1}). However by a similar
  argument as presented for the exterior derivative of $\ox_i$ one can
  see that any non-vanishing such coefficient will produce a nonzero
  derivative of $\ax_a$ or/and $\bx^A$ contradicting (\ref{dax}). From
  this one concludes that the only solution of (\ref{dox1}) is
  (\ref{dox}).}

Eqs.~(\ref{dax}) and~(\ref{dox}) imply, just as we anticipated above,
that neither $\omega_i$ nor $\tox^i$ are harmonic. In particular,
$\omega_i$ are no longer closed while the dual forms $\tox^i$  are no
longer coclosed, since at least one linear combination $e_i\tox^i$ is
exact. However, assuming for instance that $e_1$ is non-zero, the
linear combinations 
\begin{equation}
  \label{ox'}
  \ox_i' = \ox_i - \frac{e_i}{e_1}\, \ox_1\ , \qquad i\ne 1\ ,
\end{equation}
are harmonic in that they satisfy
\begin{equation}
  \dd \ox_i' = \dd^\dagger \ox_i' = 0 \ ,
\end{equation}
where we used 
$ \dd^\dagger \ox_i' = *\dd * \ox_i' \sim *\dd \tilde\ox^{\prime i}$. 
Thus there are still at least $h^{(1,1)}(Y)-1$ harmonic forms $\ox_i'$
on $\hat Y$. 
The same argument can be repeated for $H^3$ where one
finds $2h^{(1,2)}$ harmonic forms or in other words the dimension
of $H^3$ has changed by two and we have together
\begin{equation}
h^{(2)}(\hat Y) = h^{(1,1)}(Y)-1\ , \qquad
h^{(3)}(\hat Y) = h^{(3)}(Y)-2\ .
\end{equation}
Physically this can be
understood from the fact that some of the scalar fields gain a mass
proportional to the flux parameters and no longer appear as zero modes
of the compactification. Similarly, from mirror symmetry
we do not expect the occurrence of new 
zero modes on $\hat Y$ as these would correspond to additional new
massless fields in the effective action. 
This is also consistent with our expectation
that $\hat Y$ is topologically  different from $Y$ which 
stresses the
point that $Y$ and $\hat{Y}$ can only be locally close to each other in
the large complex structure limit. 

Simply from the moduli space of $\SU(3)$-structure of $\hat{Y}_\ff$
and the relation~\eqref{F4} we have conjectured the existence of a set
of forms on $\hat{Y}_\ff$ satisfying the conditions~\eqref{dax}
and~\eqref{dox} which essentially encode information about the
topology of $\hat{Y}_\ff$. We should now see if this is compatible
with a half-flat structure. In particular we find, given~\eqref{Oxz2}, 
\begin{equation}
\label{eq:dJOe}
\begin{aligned}
   \dd J &= v^i e_i \beta^0 \ , \\
   \dd \Omega &= e_i \tox^i \ . 
\end{aligned}
\end{equation}
From the standard $\SU(3)$ relation $J\wedge\Omega=0$ we have
that $\omega_i\wedge\alpha^A=\omega_i\wedge\beta^A=0$ for all $A$ and $i$
and hence in particular $J\wedge\dd J=0$. Furthermore, since the $e_i$ are
real, $\dd\Omega^-=0$. Thus we see that 
\eqref{dax} and~\eqref{dox} are consistent
with half-flat structure.\footnote{
  It would be interesting to calculate the moduli space of half-flat
  metrics on $\hat{Y}_\ff$ directly and see that it agreed with, or
  at least had a subspace, of the form given by~\eqref{Oxz2}
  and~\eqref{*J} together with~\eqref{dax} and~\eqref{dox}.}
Furthermore, since $\dd J$ and $\dd\Omega$
completely determine the intrinsic torsion $T^0$, we see that all the
components of $T^0$ are given in terms of the constants $e_i$ without
the need for any additional information.

Let us summarize. We proposed a set of rules for
identifying the light modes for compactification on $\hat{Y}$
compatible with mirror symmetry and half-flatness. We first argued
that in the limit of large complex
structure the torsion of $\hat{Y}$ is
small, and locally $\hat{Y}$ and $Y$ are metrically equivalent, even though
globally they have different topology. 
In this limit,
the light spectrum corresponds to modes
on $\hat{Y}$ which locally map to the zero modes of $Y$. 
This was made more precise by
first noting that mirror symmetry implies a one-to-one correspondence
between each pair of a
Calabi--Yau manifold $Y$ and flux $\ff\in H^4(Y,\bbZ)$ and 
a unique half-flat manifold $\hat{Y}_\ff$. 
As a consequence
the moduli space of half-flat metrics on $\hat{Y}_\ff$  has to be identical 
with the moduli space of Calabi--Yau metrics on $Y$.
In addition, the metrics on these moduli spaces agree and a basis of
forms for $J$ and $\Omega$ exist on $\hat{Y}$ which coincides with the 
corresponding basis of harmonic forms on $Y$ in the small torsion limit.
Identifying the missing NS flux $e_i$ as 
$F\sim\dd\Omega^{2,2}\sim e_i \tilde\omega^i$ 
led to a set of differential
relations  among this basis of forms in terms of the
$h^{(1,1)}(Y)$ flux parameters $e_i$. 
We further showed that these relation are compatible with the
conditions of half-flatness. 
As we will see more explicitly 
in the next section these forms give the correct basis
for expanding the ten-dimensional fields on $\hat Y$ and obtaining
a mirror symmetric effective action. We will find that the masses of the light
modes are proportional to the fluxes and thus to the intrinsic torsion
of $\hat Y$.


\subsection{The effective action}
\label{LLEA}

In this section we present the derivation of the low-energy effective action
of type IIA supergravity compactified on the manifold $\hat Y$ described in
sections \ref{sec:hf} and \ref{spectrum}.
As argued in the previous section we insist on keeping the same light spectrum
as for
Calabi--Yau compactifications and therefore the KK-reduction is closely
related to the reduction on Calabi--Yau manifolds which we recall
in appendix \ref{IIBnsflux}. The difference is that the differential forms we
expand in are no longer harmonic but instead obey
\begin{equation}
  \label{dab}
  \dd \ax_0 = e_i \tox^i\ , \qquad \dd \ax_a = \dd \bx^A = 0 \ ,\qquad
 \dd \ox_i = e_i \bx^0\ ,  \qquad \qquad \dd \tox^i = 0 \ .
\end{equation}
However we continue to demand that these forms have identical
intersection numbers as on the Calabi--Yau or in other words
obey unmodified (\ref{normYhat}).
As we are going to see shortly the relations (\ref{dab})
are responsible for generating mass terms in the effective action
consistent with the discussion in the previous section.\footnote{%
  Note that we are not expanding in the harmonic forms $\ox'_i$
  defined in (\ref{ox'}) but continue to use the non-harmonic
  $\ox_i$. The reason is that in the  $\ox_i$-basis mirror symmetry
  will be manifest. An expansion in the $\ox'_i$-basis merely
  corresponds to field redefinition in the effective action as they
  are just linear combinations of the $\ox_i$.}

Let us start  from the type IIA action in $D=10$ \cite{JP}
\begin{eqnarray}
  \label{SIIA}
  S & = & \int \, e^{-2\hat\phi} \left( -\frac12 \hat R *\! {\bf 1} + 2
  \dd \hat\phi \wg * \dd \hat\phi - \frac14  \hat H_3\wg * \hat H_3 \right)
  \nn \\ 
  & & - \frac12  \int \left(\hat  F_2 \wg * \hat F_2 + \hat F_4 \wg * \hat F_4
  \right) + \frac12 \int \hat H_3 \wedge \hat C_3 \wedge \dd \hat C_3 \, ,
\end{eqnarray}
where the notation is explained in more detail in appendix \ref{IIAcomp}.
In the KK-reduction the ten-dimensional (hatted) fields are expanded
in terms of the forms $\ox_i, \ax_A,\bx^A$ introduced in
(\ref{Oxz2})
\begin{eqnarray}
  \label{fexp}
 \hat \phi & = & \phi \ ,\qquad   \hat A_1  =  A^0 \ ,\qquad
\hat B_2 =  B_2 + b^i \ox_i      \nn \\
  \hat C_3 & = & C_3 + A^i \wg \ox_i + \xi^A \ax_A + \txi_A \bx^A \ ,
\end{eqnarray}
where $A^0,A^i$ are one-forms in $D=4$ (they will populate $h^{(1,1)}$
vector multiplets and contribute the graviphoton to the gravitational
multiplet) while $\xi^A,\txi_A,b^i$ are scalar fields in $D=4$. The
$b^i$ combine with the K\"ahler deformations $v^i$ of (\ref{Oxz2}) to form
the complex scalars $t^i = b^i + iv^i$ sitting in the
$h^{(1,1)}$ vector multiplets. The $\xi^a,\txi_a$ together
with the complex structure deformations $z^a$ of (\ref{Oxz2}) are members
of $h^{(1,2)}$ hypermultiplets  while $\xi^0,\txi_0$ together with
the dilaton $\phi$ and $B_2$ form the tensor multiplet.

The difference with Calabi--Yau compactifications results from the fact
that the derivatives of $\hat B_2, \hat C_3 $ in (\ref{fexp}) are modified
as a consequence of  (\ref{dab}) and we find
\begin{eqnarray}
  \label{dfexp}
  \dd \hat C_3 & = & \dd C_3 + (\dd A^i) \wg \ox_i
  + (\dd \xi^A) \ax_A + (\dd\txi_a) \bx^a + (\dd \txi_0 - e_i A^i)  \bx^0
  + \xi^0 e_i \tox^i \ ,\nn \\
  \dd \hat B_2 & = & \dd B_2 + (\dd b^i) \ox_i + e_i b^i \bx^0\ .
\end{eqnarray}
We already see that the scalar $\txi_0$ becomes charged precisely
due to (\ref{dab}) which is exactly what we expect from the type IIB
action. However, on the type
IIB side we have $(h^{(1,2)} +1)$ electric flux parameters while in
(\ref{dfexp}) only $h^{(1,1)}$ fluxes $e_i$ appear. The missing flux arises
from  the NS 3-form
field strength $\hat H_3 = \dd \hat B_2$ in the direction of $\bx^0$.
Turning on this additional NS flux
amounts to a shift
\begin{equation}
  \label{flux}
  \hat H_3 \to \hat H_3 + e_0 \bx^0 \, ,
\end{equation}
where $e_0$ is the additional mass parameter.
Using (\ref{dfexp}), (\ref{flux}) and
(\ref{HFdef}) we see
that the parameter $e_0$ introduced in this way naturally combines with the
other fluxes $e_i$ into
\begin{eqnarray}
  \label{modFS}
  \hat H_3 & = &  \dd B_2 + \dd b^i \ox_i + (e_i b^i +e_0) \bx^0\ ,  \\
  \hat F_4 & = & (\dd C_3 - A^0 \wg \dd B_2) + (\dd A^i - A^0 \dd b^i) \wg
  \ox_i + D \xi^A \ax_A + D \txi_A \bx^A  + \xi^0 e_i \tox^i\ ,\nn
\end{eqnarray}
where the covariant derivatives are given by
\begin{equation}
  \label{cd}
  D \txi_0 = \dd \txi_0 - e_i (A^i + b^i A^0) - e_0 A^0 , \qquad
  D \xi^A = \dd \xi^A , \qquad D \txi_a = \dd \txi_a .
\end{equation}
This formula is one of the major consequences of compactifying on $\hat Y$ (in
particular of expanding the ten-dimensional fields in forms which are not
harmonic) as one of the scalars, $\txi_0$, becomes charged.

{}From here  on the
compactification proceeds as in the massless case by inserting (\ref{modFS})
into the action (\ref{SIIA}). Except for few differences which we 
point out, the calculation continues as in appendix \ref{IIAcomp} and we are
not going to repeat this calculation here.
Using (\ref{fexpA}),  (\ref{dfexp}) and (\ref{modFS})
one can see that the parameters $e_0$ and $e_i$ give rise to new interactions
coming from the topological term in (\ref{SIIA})
\begin{eqnarray}
  \label{ltop}
  \frac12 \int_{\hat Y} \hat H_3 \wedge \hat C_3 \wedge \dd \hat C_3
  & = &  \frac{\xi^0}{2} \dd B_2 \wg A^i e_i - \frac12 \dd B_2
  \wg \left(\xi^0 (\dd \txi_0 -e_i A^i) + \xi^a \dd \txi_a - \txi_A \dd \xi^A
  \right)  \nn \\
  & & + \frac{\xi^0}{2} e_i \dd b^i \wg C_3 +\frac12 \dd b^i \wg A^j \wg \dd
  A^k \cK_{ijk}  \\
  & & - \frac{\xi^0}{2} (e_i b^i + e_0) \dd C_3 -
  \frac12(e_i b^i + e_0) \wg C_3 \wg \dd \xi^0\ ,\quad \nn
\end{eqnarray}
where $\cK_{ijk} $ is defined in (\ref{K}).

The 3-form $C_3$ in 4 dimensions carries no physical degrees of
freedom. Nevertheless 
it can not be neglected as it may introduce a cosmological constant. Moreover
when such a form interacts non-trivially with the other fields present in the
theory as in (\ref{ltop}) its dualization to a constant requires more
care. Collecting all terms which contain $C_3$ we find
\begin{equation}
  \label{SC3}
  S_{C_3} = - \frac{\cK}{2} (\dd C_3 - A^0 \wg \dd B_2) \wg * (\dd C_3 - A^0
  \wg \dd B_2) - \xi^0 \, (e_i b^i + e_0) \dd C_3 \, .
\end{equation}
As shown in \cite{BGG} the proper way of performing this dualization is by
adding a Lagrange multiplier $\lx \dd C_3$.
The 3-form $C_3$ is dual to the constant $\lx$ which was shown 
to be mirror symmetric to a RR-flux in ref.\ \cite{LM2}   and
consequently plays no role in the analysis here. 
Solving for $\dd C_3$, inserting the
result back into (\ref{SC3}) and in the end setting $\lx= 0$ we obtain the
action dual to (\ref{SC3})
\begin{equation}
  \label{Cdual}
  S_{dual} = - \frac{(\xi^0)^2}{2 \cK} (e_i b^i + e_0)^2 - \xi^0 \, (e_i b^i +
  e_0)  A^0 \wg \dd B_2 \, .
\end{equation}

Finally, in order to obtain the usual $N=2$ spectrum we dualize $B_2$ to a
scalar field denoted by $a$.
Due to the Green-Schwarz type interaction of $B_2$ (the first term in
(\ref{ltop}) and the second term in (\ref{Cdual})) $a$ is charged, but beside
that the dualization proceeds as usual.
Putting together all the pieces and after going to the Einstein frame one can
write the compactified action in the standard $N=2$ form
\begin{eqnarray}
  \label{S4A}
  S_{IIA} & = & \int \Big[ -\frac12 R *\! {\bf 1} - g_{ij} \dd t^i \wg * \dd
  {\bar t}^j - h_{uv} Dq^u \wg * Dq ^v \nn \\
  & & \qquad + \frac{1}{2}\, \IM \cN_{IJ} F^I\wg * F^{J}
  + \frac{1}{2} \, \RE \cN_{IJ} F^I \wg F^J - V_{IIA} *\! {\bf 1} \Big] ,
\end{eqnarray}
where the gauge coupling matrix $\cN_{IJ}$ and the metrics $g_{ij}, h_{uv}$
are given in (\ref{eq:N}), (\ref{gH11}) and (\ref{qktNS}) respectively.
As explained in appendix \ref{IIAcomp} the gauge couplings can be properly
identified after redefining the gauge fields $A^i \to A^i - b^i A^0$. We have
also introduced the notation $I=(0,i) = 0, \ldots, h^{(1,1)}$ and so
$A^I = (A^0, A^i)$.
Among the covariant derivatives of the hypermultiplet scalars $Dq^u$
the only non-trivial ones are
\begin{equation}
  \label{cds}
  D a = \dd a - \xi^0 e_I A^I \, ; \qquad D \txi_0 = \dd \txi_0 - e_I A^I \, .
\end{equation}
We see that two scalars are charged under a Peccei-Quinn symmetry
as a consequence of the non-zero $e_I$.

Before discussing the potential  $V_{IIA}$ let us note that the action
(\ref{S4A}) already has the form expected from the mirror symmetric
action given in appendix~\ref{IIBNSflux}. In
particular the forms $\ax_0$ and
$\bx^0$ in (\ref{dab}) single out the two scalars $\xi^0,\txi_0$
from the expansion of
$\hat C_3$. $\xi^0$ maps under mirror symmetry to
the RR scalar $l$ which is already present in the $D=10$ type IIB theory
while  $\txi_0$ maps to the charged RR scalar in type IIB.
Moreover, using these identifications one observes that the gauging
(\ref{cds})  is precisely what one obtains in the
type IIB case with NS electric fluxes turned on (\ref{action5}).

Finally, we  need to check that the potential from (\ref{S4A})
coincides with the one
obtained in the type IIB case (\ref{potIIB}).
In the case of type IIA compactified on $\hat Y$ one can identify four
distinct contributions to the potential: from the kinetic terms of $\hat B_2$
and $\hat C_3$, from the dualization of $C_3$ in 4 dimensions and
from the Ricci scalar of $\hat Y$. We study these contributions in turn.
We go directly to the four-dimensional Einstein frame which amounts to
multiplying every term in the potential by a factor $e^{4 \phi}$ coming from
the rescaling of $\sqrt {-g}$, $\phi$ being the four-dimensional
dilaton which is related to the ten-dimensional dilaton
$\hat \phi$ by $e^{-2 \phi} = e^{-2 \hat \phi} \cK $.

Using (\ref{modFS}) we see that the kinetic
term of $\hat B_2$ in (\ref{SIIA})  contributes to the potential
\begin{equation}
  \label{V1}
  V_1 = \frac{e^{2 \phi}}{4 \cK} (e_i b^i + e_0)^2 \int_{\hat Y} \bx^0 \wg *
  \bx^0 = - \frac{e^{- 2 \phi}}{4 \cK} (e_i b^i + e_0)^2 \left[(\IM
    \cM)^{-1}\right]^{00} \, ,
\end{equation}
where the integral over $\hat Y$ was performed using (\ref{star}), (\ref{A-N})
and (\ref{oxstar}).
Similarly, the kinetic term of $\hat C_3$ produces the following piece in the
potential
\begin{equation}
  \label{V2}
  V_2 = e^{4 \phi} \, \frac{(\xi^0)^2}{8 \cK}\, e_i e_j g^{ij}\ ,
\end{equation}
where $g^{ij}$ arises after integrating over $\hat Y$ using (\ref{oxstar}).
Furthermore, (\ref{Cdual}) contributes
\begin{equation}
  \label{V3}
  V_3 = e^{4 \phi} \, \frac{(\xi^0)^2}{2 \cK} (e_i b^i + e_0)^2 \ .
\end{equation}
Combining (\ref{V1}), (\ref{V2}) and (\ref{V3}) we arrive at
\begin{eqnarray}
  \label{potA}
    V_{IIA} & = &V_g + V_1 + V_2 + V_3 \\
    & =& V_g - \frac{e^{2 \phi}}{4 \cK} (e_i b^i + e_0)^2
    \left[(\IM \cM)^{-1}\right]^{00} - e^{4 \phi} \, \frac{(\xi^0)^2}{2}
    e_I e_J \left[(\IM \cN)^{-1} \right]^{IJ}\ ,\nn
\end{eqnarray}
where we used
the form of the matrix $(\IM \cN)^{-1}$ given in (\ref{ImN-1e}).
$V_g$ is a further contribution to the potential which arises from
the Ricci scalar. Since $\hat Y$ is no longer Ricci-flat $R$
contributes to the potential and in this way provides
 another sensitive test of the half-flat geometry.

In appendix \ref{Rhf} we show that for half-flat manifolds
the Ricci scalar  can be written in terms of the contorsion as
\begin{equation}
  \label{R}
  R = - \CT_{mnp} \CT^{npm} - \frac{1}{2} \epsilon^{mnpqrs} (\nabla_m \CT_{npq} -
  \CT_{mp}{}^l \CT_{nlq}) J_{rs} \ ,
\end{equation}
which, as expected, vanishes for $\CT=0$.
In order to evaluate the above expression we first we have to give a
prescription about how to compute $\nabla_m \CT_{npq}$. Taking into
account that at in the end the potential in the four-dimensional theory appears
after integrating over the internal manifold $\hat Y$
we can integrate by parts and `move' the covariant derivative to act
on $J$. This in turn can be computed by using the fact that $J$ is covariantly
constant with respect to the connection with torsion (\ref{torJO}).
Replacing the contorsion $\CT$ from (\ref{T-K}), going to complex indices and
using the defining relations for the torsion (\ref{su3Nt}), (\ref{T3}) and 
(\ref{F4}) one can find after some straightforward but tedious algebra
the expression for the Ricci scalar.
The calculation is presented in appendix \ref{Rhf} and here we only record the
final result
\begin{equation}
  \label{Rf}
  R = - \frac{1}{8} \, e_i e_j g^{ij} \left[(\IM \cM)^{-1}
  \right]^{00} \ .
\end{equation}
Taking into account the factor $\frac{e^{-2\hat \phi}}{2}$ which
multiplies the Ricci scalar in the ten-dimensional action (\ref{SIIA}) and the
factor $e^{4 \phi}$ coming from the four-dimensional Weyl rescaling one obtains
the contribution to the potential coming from the gravity sector to be
\begin{equation}
  \label{Vg}
  V_g = - \frac{e^{2 \phi}}{16 \cK} \, e_i e_j g^{ij} \left[(\IM
  \cM)^{-1}\right]^{00}\ .
\end{equation}
Inserted into (\ref{potA}) and using again (\ref{ImN-1e}) we can finally
write the entire potential
which appears in the compactification of type IIA supergravity on $\hat Y$
\begin{equation}
  \label{potAfin}
  V_{IIA} = - \frac{e^{4 \phi}}{2} \left( (\xi^0)^2 - \frac{e^{-2 \phi}}{2}
  \left[(\IM \cM)^{-1}\right]^{00}  \right) e_I e_J
  \left[(\IM \cN)^{-1} \right]^{IJ} .
\end{equation}

In order to compare this potential to the one obtained in type IIB case
(\ref{potIIB}) we should first see how the formula (\ref{potAfin}) changes under
the mirror map. We know that under mirror symmetry the gauge coupling matrices
$\cM$ and $\cN$ are mapped into one another. In particular this means
that\footnote{In order to avoid confusions we have added the label $A/B$ to
  specify the fact that the corresponding quantity appears in type IIA/IIB
  theory.} 
\begin{equation}
  \left[(\IM \cM_A)^{-1}\right]^{00} \leftrightarrow \left[(\IM
  \cN_B)^{-1} \right]^{00} = - \frac{1}{\cK_B} \ .
\end{equation}
where we used the expression for $(\mathrm{Im} \; \cN)^{-1}$ from (\ref{ImN-1e}).
With this observation it can be easily seen that the type IIA potential
(\ref{potAfin}) is precisely mapped into the type IIB one (\ref{potIIB})
provided one identifies the electric flux parameters
$e_I \leftrightarrow \q_A$
and the four-dimensional dilatons on the two sides.

To summarize the results obtained in this section, we have seen that the 
low-energy effective action of type IIA theory compactified on $\hat Y$ is
precisely the mirror of the effective action obtained in appendix
\ref{IIBNSflux} for type IIB theory compactified on $Y$ in the presence of NS
electric fluxes. This is our final argument that the half-flat manifold
$\hat Y$ is the right compactification manifold for obtaining the mirror
partners of the NS electric fluxes of type IIB theory. In particular the
interplay between the gravity and the matter sector which resulted in the
potential (\ref{potAfin}) provided  a highly nontrivial check of
this assumption.

\section{Conclusions}
\label{conc}\setcounter{equation}{0}

In this paper we propose that type IIB (or alternatively IIA)
compactified on a Calabi--Yau threefold $\tilde{Y}$ with electric NS
three-form flux is mirror symmetric to type IIA (respectively IIB)
compactified on a half-flat manifold $\hat{Y}$ with $SU(3)$ structure. 
The manifold $\hat{Y}$ is neither complex nor is it Ricci-flat. 
Nonetheless, though topologically distinct, it is closely related to
the ordinary Calabi--Yau mirror partner $Y$ of the original threefold
$\tilde{Y}$.  
In particular, we argued that the moduli space of half-flat metrics on
$\hat{Y}$ must be the same as the moduli space of Calabi--Yau metrics
on $Y$.  
Furthermore, it is the topology of $\hat{Y}$ that encodes the
even-dimensional NS-flux mirror to the original $H_3$-flux on
$\tilde{Y}$.  

We established this correspondence first by considering toroidal and
orbifold compactification of type IIB where the mirror map is realized
as a T-duality transformation and therefore can be performed
explicitly. 
Similarly, in the SYZ picture, where mirror Calabi--Yau
threefolds are viewed as special Lagrangian $T^3$-fibrations, in the
large complex structure limit, the mirror map is again realized as a 
simple T-duality. In both cases, starting with a Calabi--Yau background
with NS three-form flux, the mirror configuration is purely
geometrical, with no $H_3$-flux and trivial dilaton, and the resulting
geometry has $\SU(3)$-structure and satisfies the half-flat
conditions~(\ref{eq:hf}).

We further strengthened this proposal by deriving the low-energy
type IIA effective action in the supergravity limit
and showing that it is exactly equivalent to the appropriate type IIB
effective action. In particular, the resulting potential delicately
depends on the non-vanishing Ricci scalar of the half-flat geometry
and thus provided a highly non-trivial check on our proposal.

It is interesting to note that one particular NS flux $e_0$ played a
special role in that it did not arise from the half-flat geometry but,
as in type IIB, appeared as a NS three-form flux $H_3\in
H^{(3,0)}(\tilde{Y})$. In this context, it appears that mirror
symmetry only acts on the `interior' of the Hodge diamond in that it
exchanges $H^{(1,1)} \leftrightarrow H^{(1,2)}$ but
leaves $H^{(3,3)}\oplus H^{(0,0)}$ and $H^{(3,0)}\oplus H^{(0,3)}$
untouched. Put another way, it appears that it is the same single NS
electric flux which is associated to both  $H^{(3,0)}\oplus
H^{(0,3)}$ and $H^{(3,3)}\oplus H^{(0,0)}$ on a given Calabi--Yau
manifold.

We found that requirements of mirror symmetry provided a number of
conjectures about the geometry of the half-flat manifold
$\hat{Y}$. For instance the cohomology groups of $\hat Y$ shrink
compared to those of $Y$ in that the Hodge numbers $h^{(1,1)}$ and
$h^{(1,2)}$ are reduced by one.\footnote{%
This fact could be related to a possible generalization of the
 proposal by M.~Reid \cite{reid} where similarly the Hodge numbers
 decrease due to a resolution of small three-cycles as two-cycles
 such that the manifold after the `topology-changing' transition is
 still complex but non-K\"ahler. (In our terminology these manifolds
 correspond to $\W_1=\W_2=0$.)} 
In addition, a non-standard KK reduction had to be performed in order
to obtain masses for some of the scalar fields. This in turn led us to
make a number of assumptions which need
to be better understood from a mathematical point of view.
One particular conjecture is the following. In general, the electric
NS $H_3$-flux maps under mirror symmetry to some element $\zeta\in
H^4(Y)$. Mirror symmetry would appear to imply that 
\begin{quote}
   for all integer fluxes $\zeta\in H^4(Y,\bbZ)$ there should be
   a unique manifold $\hat{Y}_\zeta$ admitting a family of half-flat
   metrics such that the moduli space of such metrics $\mathcal{M}(Y_\zeta)$
   is equal to the moduli space $\mathcal{M}(Y)$ of Calabi--Yau metrics on $Y$.  
\end{quote}
In the SYZ picture we expect that all these manifolds $\hat{Y}_\zeta$
and $Y$ appear locally as $T^3$-fibrations over the same
base. However, the fibration of $\hat{Y}_\zeta$ is generalized so that
the total space is no longer Calabi--Yau. 
We note that it should be possible to determine 
this moduli space of half-flat geometries directly from its
definition and without relying on the physical relation
with Calabi--Yau threefold compactification.\footnote{In this respect
a generalization of ref.~\cite{H2} might be useful.}
Moreover, a more precise mathematical statement about the relationship
between a given Calabi--Yau threefold $Y$ and its `cousin'
half-flat geometry on $\hat{Y}$ should also be possible. 
Finally, our analysis only treated electric NS fluxes. The
discussion of the magnetic ones is technically more involved as on the type
IIB side a massive RR two-form appears which has no obvious
counterpart on the type IIA side.
We hope to report on all of these issues in the near future \cite{GLMW}.

The relevance of half-flat geometries can also be understood
from a different point of view. The four-dimensional type IIB effective
action with NS background fluxes admits $N=1$ BPS domain-wall
solutions \cite{BCL}. In the mirror symmetric type IIA action these
domain-walls have to be entirely geometrical with no fluxes turned on.
Indeed, as shown in refs.~\cite{H,CS}
half-flat manifolds when appropriately
fibered over an interval always admit a metric of $G_2$ holonomy.
It is precisely this geometry which governs the three-dimensional $N=1$
effective action on the domain-wall.
This is closely related to the discussion in~\cite{AMV}. There it was
shown that starting from a type IIB theory with both RR- and NS
background fluxes the conjectured mirror symmetric type IIA theory is
related to a similarly purely geometrical compactification of M-theory
on a $G_2$ manifold.  

The $N=1$ BPS domain walls
can be conveniently characterized by a holomorphic superpotential 
$W$ \cite{GVW,gukov} which on the type IIB side takes the form
\begin{equation}
W_B = \int_{\tilde Y} \Omega\wedge (F_3 + \tau H_3) \ ,
\end{equation}
where $\tau = l+ ie^{-\phi}$ is the complex type IIB dilaton.
For $H_3=0$ the type IIA mirror superpotential is given by \cite{GVW,gukov}
\begin{equation}
W_A^{RR} = \int_Y F_6 +  J\wedge F_4 + J^2\wedge F_2 + J^3 F_0 \ .
\end{equation}
For $H_3\neq 0$ the type IIA mirror $W$ should have the exact
same structure but with complexified fluxes \cite{Vafa}.
In this paper we only considered electric fluxes and hence
we only discovered the real four-form $F\sim \dd\Omega_+$ in 
\eqref{F22} and \eqref{F4}. Hence the corresponding superpotential
should have the form, 
\begin{equation}
   W_A^{NS} = \int_{\hat Y} (J+\ii B) \wedge 
        (\dd C_3 + \ii\ee^{-\phi}\dd\Omega_+) \ .
\end{equation}
(Recall that the IIA analog of $l$ comes from part of the $C_3$ light
modes, hence the $\dd C_3$ term.) Indeed, it is easy to check that,
truncating the theory so we consider only the NS zero modes, the
relevant four-dimensional $N=1$ action with such a superpotential does
admit appropriate BPS domain walls solutions.

{}From this perspective it is easy to conjecture that the magnetic
fluxes introduce a non-trivial two form with $W_A^{NS} \sim \int_{\hat
  Y} J^2\wedge \dd^\dagger\Omega^+$ being a natural candidate. Note
that, this is equivalent, given the action of the Hodge star operator,
to a term $\int_{\hat{Y}}J\wedge\dd\Omega^-$. This would suggest that
magnetic fluxes require a further generalization of the half-flat
geometry allowing the possibility of $\dd\Omega^-\neq 0$. In general,
for an effective action with $N=2$ supersymmetry all that is required
is that one compactifies on a manifold with $\SU(3)$ structure. 

We can also make some speculations on the relation of the half-flat
geometry of $\hat{Y}$ to the Calabi--Yau manifold $Y$. Recall that 
type IIB theory has an S-duality symmetry exchanging RR and NS
three-form flux. The IIA theory must have the same symmetry in
four dimensions. Consider first the electric RR flux. $F_4$ is then a
non-zero harmonic form so that we only have $F_4=\dd C_3$  
locally. (Mathematically $C_3$ is the connection on a ``gerbe'', a
sort of generalization of a bundle, as described for instance in~\cite{gerbe}.)
Without flux, the moduli of $C_3$ appear in hypermultiplets 
with the complex structure moduli of $\Omega$ on $Y$, and are paired
by the S-duality. In other words the S-dual of $F_4$ flux would appear
to be that statement that $\dd\Omega$ is now non-zero. (Note exactly
the same kind of argument was made earlier relating $\dd B_2$ and
$\dd\Omega_+$ by mirror symmetry.) If one really takes the analogy
seriously not only should the flux $\dd\Omega$ be non-zero but
$\Omega$ should only be defined locally on $Y$. In other words, it
appears that the mirror configuration is the same three-fold $Y$ but
now, since $\Omega$ is not globally defined, we now longer have an
$\SU(3)$ structure. However, this is, in fact, not the situation. The
actual mirror configuation as we have seen, is to take a 
different manifold $\hat{Y}$, in some sense a twisted version of $Y$,
on which $\Omega$ \textit{is} globally defined. Essentially we
exchange $Y$ with a non-trivial gerbe for a new manifold $\hat{Y}$
with a trivial gerbe. On $Y$ the four-form $F_4$ is actually dual to
a two-form, which in turn is just the field strength of a conventional
$U(1)$ bundle. From this point of view the twisting of $\Omega$ on $Y$ may
simply be encoded by a some $U(1)$ bundle, perhaps with a gauge action
on the phaze of $\Omega$. From this perspective the new manfold $Y$
might be defined by something like the space of covariantly constant
sections (in general multiple covers of $Y$) in this bundle. Such a
construction might then allow us to define the basis of forms on
$\hat{Y}$ introduced in~\ref{spectrum} in terms of a twisted
cohomology on $Y$. 

Let us end by noting that the ideas in this paper apply to a number of
other situations. In general, it appears that whenever one considers
compactifications on some supersymmetric manifold $Z$ with non-trivial
flux, one should at the same level allow for compactifications on
a generalized manifold $\hat{Z}$. To preserve a supersymmetric
effective action, $\hat{Z}$ should have at least the same
$G$-structure as $Z$, but it need no longer have special holonomy. It
would be nice to know in general what conditions one must impose on
$\hat{Z}$. However, just from the current work a number of
possibilities can be considered. First, since we are considering
NS-fluxes, it is natural to take the conjectures of this paper
directly over to type I and heterotic theories. Thus, with trivial
gauge fields, half-flat compactifications of the heterotic string
should be dual to compactifications with electric
$H_3$-flux. Similarly type II compactifications on $G_2$-holonomy
manifolds with $H_3$-flux should be dual to compactification on a
particular seven-manifold $\hat{Z}$ with $G_2$ structure. Since the
manifolds with flux admit BPS domain walls, $\hat{Z}$ fibered over an
interval should be a manifold with $\Spin(7)$-holonomy. Using the
results of~\cite{H}, this implies that $\hat{Z}$ has
``co-calibrated'' $G_2$ structure~\cite{CS}.


\vspace{1cm}
\appendix
\noindent
{\Large {\bf Appendix}}
\renewcommand{\theequation}{\Alph{section}.\arabic{equation}}

\setcounter{equation}{0}\setcounter{section}{0}

\section{Conventions and notations}
\label{conv}

Throughout the paper we use the conventions from \cite{LM2} (see appendix A of
this paper). Beside this we use the following conventions.
\begin{itemize}
\item Indices $m,\, n, \, p, \ldots = 1, \ldots, 6$ label real internal
  coordinates. When we use complex coordinates we label them with
  $\ax,\bx = 1,2,3,\  \ab, \bb =1,2,3$.
\item The Riemann curvature tensor is defined as
  \begin{equation}
    \label{RTexp}
    R_{mnp}{}^q = \partial_m \phi_{np}{}^q - \partial_n \phi_{mp}{}^q
    - \phi_{mp}{}^r \phi_{nr}{}^q + \phi_{np}{}^r \phi_{mr}{}^q,
  \end{equation}
  where $\phi$ denotes a general connection that contains two contributions
$\phi_{mn}{}^p = \Gx_{mn}{}^p + \CT_{mn}{}^p$ where
$\Gx_{mn}{}^p=\Gx_{nm}{}^p$
denote the Christoffel symbols and $\CT_{mn}{}^p$ is the contorsion which
we define more precisely in appendix~\ref{acs}.
For the Ricci tensor we use $R_{np} = R_{nmp}{}^m$. (Note that differs by a
minus sign from the one used in \cite{LM2})

\item We define the $\epsilon$-symbol to be $\epsilon^{123456} = +1$.
The indices
  are lowered with the metric. It follows that in terms of `complex
  indices' one has, as a result of the $\SU(3)$ structure,
  \begin{equation}
    \epsilon^{\ax \bx \cx \ab \bb \cb} = - i \epsilon^{\ax \bx \cx} \,
    \epsilon^{\ab \bb \cb} \ .
  \end{equation}
where similarly $\epsilon^{123}=\epsilon^{\bar{1}\bar{2}\bar{3}}=+1$. 
\item For the gamma matrices we use the conventions from \cite{AvP}.
In particular the gamma matrices on the internal space are chosen to be
hermitian matrices satisfying
\begin{equation}
  \{\Gx_m, \Gx_n\} = 2 g_{mn}\ .
\end{equation}
The chirality operator $\Gx_7$ is defined as
\begin{equation}
  \Gx_7 = i \Gx_1 \ldots \Gx_6 = \frac{i}{6!} \epsilon_{m_1 \ldots m_6}
  \Gx^{m_1} \ldots \Gx^{m_6}\ .
\end{equation}
Majorana spinors on the six-dimensional internal space can be defined if we
adopt the following conventions for the charge conjugation matrix $\cC$
\begin{equation}
  \cC^T = \cC \ , \qquad \qquad \Gx_m^T = - \cC \Gx_m \cC^{-1} \ ,
\end{equation}
while the Majorana condition on a spinor $\eta$ reads
\begin{equation}
  \eta^\dagger = \eta^T \cC\ .
\end{equation}
Symmetry properties of the gamma matrices and $\cC$ with the above conventions
imply that for a commuting Majorana spinor $\eta$ the following quantities
vanish \cite{AvP}
\begin{equation}
  \label{asym}
  \eta^\dagger \Gx_{(1)} \eta = \eta^\dagger \Gx_{(2)} \eta = \eta^\dagger
  \Gx_{(5)} \eta = \eta^\dagger \Gx_{(6)} \eta = 0\ ,
\end{equation}
where by $\Gx_{(n)}$ we have denoted the antisymmetric product of $n$ gamma
matrices
\begin{equation}
  \label{Gxn}
  \Gx_{(n)} = \Gx_{m_1 \ldots m_n} = \Gx_{[m_1}  \ldots \Gx_{m_n]}\ .
\end{equation}

\end{itemize}

\section{Type II theories compactified on a Calabi--Yau 3-folds}
\label{IIBnsflux}

In this appendix we recall the known results of type II compactifications
on Calabi--Yau threefolds $Y$ in order to make the paper more self-contained
and to supplement the discussion and conventions used in section \ref{LLEA}.
In \ref{IIAcomp} we recall type IIA compactified on $Y$ without background
fluxes while in \ref{IIBNSflux} we summarize the results of type IIB
with NS three-form flux turned on.

\subsection{Type IIA compactification without fluxes}
\label{IIAcomp}
\CY{} compactifications of type IIA theory were first considered in \cite{BCF}.
We start from the following action in 10 dimensions
\begin{eqnarray}
  \label{SIIA10}
  S & = & \int \, e^{-2\hat\phi} \left( -\frac12 \hat R *\! {\bf 1} + 2
  \dd \hat\phi \wg * \dd \hat\phi - \frac14  \hat H_3\wg * \hat H_3 \right)
  \nn \\ 
  & & - \frac12  \, \int \left(\hat  F_2 \wg * \hat F_2 + \hat F_4 \wg * \hat
  F_4 \right) + \frac12 \int \hat H_3 \wedge \hat C_3 \wedge \dd \hat C_3 \, ,
\end{eqnarray}
where
\begin{equation}\label{HFdef}
\hat H_3 = \dd \hat B_2\ ,\qquad \hat F_2 = \dd \hat A_1\ ,\qquad
\hat F_4 = \dd \hat C_3 - \hat A_1 \wedge\hat H_3\ ,
\end{equation}
and $\hat \phi$ is the dilaton. (The $\hat{}$ is used to denote the fields
in $D=10$.)

In the KK-reduction we expand the
ten-dimensional fields in terms of harmonic forms on $Y$
\begin{eqnarray}
  \label{fexpA}
  \hat A_1 & = & A^0 \ ,\nn \\
  \hat C_3 & = & C_3 + A^i \wg \ox_i + \xi^A \ax_A + \txi_A \bx^A \ ,\\
  \hat B_2 & = & B_2 + b^i \ox_i \ ,\nn
\end{eqnarray}
where $C_3$ is a three-form, $B_2$ a two-form, $(A^0,A^i)$ are one-forms
and $b^i, \xi^A, \txi_A$ are scalar fields in $D=4$.
$\ox_i, \ i = 1,\ldots, h^{(1,1)}$ are harmonic $(1,1)$-forms which
form a basis of $H^{(1,1)}(Y, \mathbf{Z})$ while
$(\ax_A, \bx^A)$ are harmonic three-forms which form a real basis of
$H^3(Y, \mathbf{Z})$. They are  normalized as follows
\begin{eqnarray}
  \label{norm}
  \int_{Y} \ax_A \wg \bx^B & =& \dx_A^B\ =\
  -\int_{Y} \bx^B \wg \ax_A \ , \qquad A,B = 0,\ldots, h^{(1,2)}\ ,\nn\\
  \int_{Y} \ax_A \wg \ax_B &=& \int_{Y} \bx^A \wg \bx^B\ =\ 0\ .
\end{eqnarray}
Furthermore the deformations of the \CY{} metric can be divided into two classes,
K\"ahler class and complex structure deformations, each producing a set of
scalar fields (moduli). The K\"ahler class moduli $v^i$ are real and in
one to one correspondence with the elements of $H^{(1,1)}(Y, \mathbf{Z})$
while the complex structure moduli, $z^a, a = 1,\ldots, h^{(1,2)}$
are complex and are counted by the
elements of $H^{(2,1)}(Y, \mathbf{Z})$. The K\"ahler class
moduli are combined with the scalars $b^i$ defined in  (\ref{fexpA})
into  complex scalar fields $t^i = b^i + i v^i$.
These fields together with the one-forms $A^i$ defined in (\ref{fexpA})
combine into $h^{(1,1)}$ vector multiplets $(A^i,t^i)$.
The $\xi^a,\txi_a$ together
with the complex structure deformations $z^a$ are members
of $h^{(1,2)}$ hypermultiplets  while $\xi^0,\txi_0$ together with
the dilaton $\phi$ and $B_2$ form the tensor multiplet.
$A^0$ in (\ref{fexpA}) is the graviphoton which
together with the four-dimensional metric $g_{\mu\nu}$ describes the bosonic
components of the gravitational multiplet.

The matter part of the four-dimensional low-energy effective action can be
obtained by replacing the expansion (\ref{fexpA}) in the ten-dimensional action
(\ref{SIIA10}) and performing the integrals over the \CY{} space. The
integrals we abbreviate as
\begin{eqnarray}
  \label{K}
  \cK & = & \frac16 \int_{Y} J \wg J \wg J  \, , \quad  \cK_i = \int_{Y}
  \ox_i \wg J \wg J \ , \\
  \cK_{ij} & = & \int_{Y} \ox^i \wg \ox ^j \wg J \ , \quad \cK_{ijk} =
  \int_{Y} \ox^i \wg \ox^j \wg \ox^k  \ ,\nn
\end{eqnarray}
where $J$ is the K\"ahler form which can be expanded in terms of the basis 
$\ox_i$  as
\begin{equation}\label{Jexp}
  J= v^i \ox_i \ .
\end{equation}
For the  gravitational moduli we here only record the 
results obtained in the literature \cite{BCF,CdO}.
The metric on the complexified K\"ahler cone is
\begin{equation}
  \label{gH11}
  g_{i j}  = \frac{1}{4 \cK} \int_{Y} \ox_i \wg * \ox_j = -\frac14
  \left(\frac{\cK_{i  j}}{\cK} -  \frac14 \, \frac{\cK_i
\cK_{j}}{\cK^2}\right)\ ,
\end{equation}
which is K\"ahler i.e.\ $g_{i j} = \del_i \bar \del_j K$
with the K\"ahler potential $K$ given by
\begin{equation}
  \label{KpotK}
  e^{-K} = 8 \cK \ .
\end{equation}

On a \CY{} threefold $H^{(2,2)} (Y)$ is dual to
$H^{(1,1)} (Y)$ and it is useful to introduce the dual basis $\tilde \ox^i$
normalized by
\begin{equation}
  \label{normH2}
  \int_{Y} \ox_i \wg \tilde \ox^j\ =\ \dx_i^j\ .
\end{equation}
With this normalization the following relations hold
\begin{equation}
  \label{oxstar}
  g^{ij} = 4 \cK \int_{Y} \tilde \ox^i \wg * \tilde \ox^j \, , \quad
  * \ox_i = 4 \cK g_{ij} \tilde \ox^j \, , \quad
  * \tilde \ox^i = \frac{1}{4 \cK}\, g^{ij} \ox_j \, ,
\quad \ox_i \wg \ox_j \sim \cK_{ijk} \tilde \ox^k \, ,
\end{equation}
where the symbol $\sim$
denotes the fact that the quantities are in the same cohomology class.

For the complex structure deformations the metric $g_{ab}$
is also K\"ahler with the K\"ahler potential given by
\begin{equation}
  \label{Kpotcs}
  e^{-K}\ =\ i \int_Y \Ox \wg \bar \Ox\ =\  \cK\, ||\Ox||^2 \ ,
\end{equation}
where $\Ox$ is the holomorphic $(3,0)$ form on the \CY{} space
and $||\Ox||^2 \equiv \frac{1}{3!} \Ox_{\ax\bx\cx}\bar \Ox^{\ax\bx\cx}$.
$\Omega$ can be expanded in terms of $(\ax_A, \bx^A)$ as
\begin{equation}
  \label{Oxz}
  \Ox = z^A\, \ax_A - \cF_A\, \bx^A\ , \quad A= 0,1,\ldots,h^{(1,2)}\ ,
\end{equation}
where $z^A=(1, z^a)$ are the deformations of the complex structure and
$\cF_A$ is the derivative of the
$N=2$ prepotential.
This geometry is defined more precisely for example in ref.\ \cite{CdO}
but for our purpose here we only need to record that $\cF_A$ is a function
of the $z^a$.

In order to evaluate the integrals in the reduction we need to recall
that the Hodge-dual basis $(* \ax_A,* \bx^A)$ is related to $(\ax_A, \bx^A)$
via
\begin{equation}
  \label{star}
  * \ax_A =  {A_A}^B \, \ax_B + B_{AB} \, \bx^B \ , \qquad
  * \bx^A = C^{AB} \, \ax_B + {D^A}_B \, \bx^B\ ,
\end{equation}
where the matrices $A, \ B, \ C$ are determined by a matrix $\cM$ according to
\cite{CDF,Suz}
\begin{eqnarray}
  \label{A-N}
  A & = & \left(\RE \cM \right) \left(\IM \cM \right)^{-1}\ , \nn \\
  B & = & - \left(\IM \cM \right) - \left(\RE \cM \right) \left(\IM
  \cM\right)^{-1} \left(\RE \cM \right)\ , \nn \\
  C & = & \left(\IM \cM \right)^{-1}\ .
\end{eqnarray}
$\cM$ in turn is determined in terms of the $N=2$ prepotential $\cF$
but we do not recall this somewhat involved relation here (see, for example,
\cite{CDF}).
We should note that $\cM$ depends non-trivially on the complex
structure moduli $z^a$ and
plays the role of the gauge couplings for the case of type IIB
theory to which we will turn shortly.

With these expressions we can reduce the different terms appearing in the
action (\ref{SIIA10})
\begin{eqnarray}
  \label{HH}
  - \frac14 \int_{Y_3}\hat H_3 \wg * \hat H_3
  & = & - \frac{\cK}{4} \, \dd B_2 \wg * \dd B_2 - \cK g_{ij} \dd b^i \wg *
  \dd b^j \
  ,\nn \\ [2mm]
  - \frac12 \int_{Y_3} \hat F_2 \wg * \hat F_2 & = & - \frac{\cK}{2} \, \dd A^0
  \wg * \dd A^0 \ ,\\ [2mm]
  - \frac12 \int_{Y_3} {\hat {F}_4} \wg * {\hat {F}_4} & = & - \frac{\cK}{2}
  \, (\dd C_3 - \dd A^0\wedge B_2) \wg * (\dd C_3 - \dd A^0\wg B_2) \nn \\
  & & - 2 \cK g_{ij} (\dd A^i - A^0 \dd b^i) \wg * (\dd A^j - A^0 \dd b^j) \nn
  \\ 
  & & + \frac{1}{2}\left(\IM \cM ^{-1} \right)^{AB}
  \Big[ \dd \tilde\xi_A +  \cM_{AC} \dd \xi^C \Big] \wg * \Big[ \dd
  \tilde\xi_B + \bar \cM_{BD} \dd \xi^D \Big] \ , \nn \\ [2mm]
  \frac12 \int_{Y} \hat H_3 \wedge \hat C_3 \wedge \dd \hat C_3
  & = &  - \frac12 \dd B_2 \wg (\xi^A \dd \txi_A - \txi_A \dd \xi^A) +\frac12
  \dd b^i 
  \wg A^j \wg \dd A^k \cK_{ijk}  \, . \nn
\end{eqnarray}

The dualization of a 3-form $C_3$ in 4 dimensions produces a
contribution to the cosmological constant. As shown in \cite{LM2}
this constant can be viewed as a specific RR-flux. Since we are not interested
in RR-fluxes here we choose it to be zero and hence  discard
the contribution of $C_3$ in 4
dimensions. Thus the only thing we still need to do in order to recover the
standard spectrum of $N=2$ supergravity in 4 dimensions is to dualize the
2-form $B_2$ to an axion $a$. The string frame action for $B_2$
\begin{equation}
  \label{actH}
  \cL_{H_3} = - \frac{1}{4} e^{-2\phi}H_3 \wg * H_3  + \frac12 H_3 \wg
  \left(\tilde\xi_A \dd \xi^A -\xi^A \dd \tilde\xi_A \right)   \, ,
\end{equation}
produces the following equation of motion for $B_2$
\begin{equation}
  \dd \left( e^{-2\phi} *\dd B_2 - \tilde\xi_A \dd \xi^A +\xi^A \dd
  \tilde\xi_A \ , \right) = 0 \ ,
\end{equation}
which can be satisfied by setting
$\dd a= \left(e^{-2\phi} * \dd B_2- \tilde\xi_A \dd \xi^A +\xi^A \dd
  \tilde\xi_A\right)$. 
The equation of motion for $a$ (derived from the Bianchi identity for $H_3$)
\begin{equation}
  \dd *\left(e^{2\phi} \dd a + \tilde\xi_A \dd \xi^A - \xi^A \dd
  \tilde\xi_A\right) = 0 \ ,
\end{equation}
can in turn be obtained from the action
\begin{equation}
  \label{dact}
  \cL_a = - \frac{e^{2\phi}}{4} \,
  \Big[\dd a + (\tilde\xi_A \dd \xi^A -\xi^A \dd \tilde\xi_A)
  \Big] \wg *  \Big[\dd a + (\tilde\xi_A \dd\xi^A-\xi^A \dd \tilde\xi_A) \Big]
  \ , 
\end{equation}
which is the dual action of (\ref{actH}).
The usual $N=2$ supergravity couplings can be read off after redefining the
gauge fields $A^i \to A^i - b^i A^0$ and introducing the collective notation
$A^I = (A^0, A^i)$ where $I=(0,i) = 0, \ldots, h^{(1,1)}$.

Collecting all terms from (\ref{HH}), (\ref{dact}) and taking into account the
scalars coming from the gravity sector \cite{BCF} and after going to the Einstein
frame the four-dimensional action becomes
\begin{eqnarray}
  \label{S4A0}
  S_{IIA} & = & \int \Big[ -\frac12 R ^* {\bf 1} - g_{ij} \dd t^i \wg * \dd
  {\bar t}^j - h_{uv} \dd q^u \wg * \dd q ^v \nn \\
  & & \qquad + \frac{1}{2}\, \IM \cN_{IJ} F^I\wg * F^{J}
  + \frac{1}{2} \, \RE \cN_{IJ} F^I \wg F^J \Big] ,
\end{eqnarray}
where the gauge coupling matrix $\cN$ has the form
\begin{eqnarray}
  \label{eq:N}
  \RE \cN_{00} = - \frac13 \cK_{ijk} b^i b^j b^k \, , & & \IM \cN_{00} = -
  \cK + \left(\cK_{ij} - \frac14 \, \frac{\cK_i \cK_j}{\cK} \right) b^i
  b^j\ ,
  \nn \\
  \RE \cN_{i0} = \frac12 \cK_{ijk} b^j b^k \ , & & \IM \cN_{i0} = -
  \left(\cK_{ij} - \frac14 \, \frac{\cK_i \cK_j}{\cK} \right) b^j \ ,\\
  \RE \cN_{ij} = - \cK_{ijk} b^k \ , & & \IM \cN_{ij} = \left(\cK_{ij} - \frac14
  \, \frac{\cK_i \cK_j}{\cK} \right) \ ,\nn
\end{eqnarray}
and $h_{uv}$ is the $\sigma$-model metric for the scalars in the
hypermultiplets \cite{FeS}
\begin{eqnarray}
  \label{qktNS}
  h_{uv} \dd q^u \wg * \dd q ^v & = &  \dd \phi \wg * \dd \phi + g_{ab} \dd
  z^a \wg * \dd \bar z^b 
  \\
  & & + \frac{e^{4\phi}}{4} \, \Big[ \dd a +
  (\tilde\xi_A \dd \xi^A-\xi^A \dd \tilde\xi_A) \Big] \wg *
  \Big[\dd a +
  (\tilde\xi_A \dd \xi^A-\xi^A \dd \tilde\xi_A) \Big] \nn \\
  & & - \frac{e^{2\phi}}{2}\left(\IM \cM^{-1} \right)^{AB}
  \Big[ \dd \tilde\xi_A + \cM_{AC} \dd \xi^C \Big]
   \wg * \Big[ \dd \tilde\xi_B + \bar \cM_{BD} \dd\xi^D \Big]  \,
  .\nn
\end{eqnarray}

In the main part of the paper we also need the form of
the inverse gauge couplings which is given by
\begin{equation}
  \label{ImN-1e}
  \left(\IM \cN\right)^{-1} = - \frac{1}{\cK}\left(
    \begin{array}{cc}
      1 & b^i \\
      & \\
      b^i \quad & \frac{g^{ij}}{4} + b^i b^j
    \end{array} \right) \, .
\end{equation}

\subsection{Type IIB theory with NS flux}
\label{IIBNSflux}

This appendix is intended to outline the
main features of the low-energy effective action of type IIB supergravity
compactified on \CY{} 3-folds in the presence of NS 3-form flux $H_3$.
For only electric NS fluxes the effective action was derived in \cite{JM,GD}
while the potential for both electric and magnetic fluxes appeared
in \cite{TV}. However, the entire bosonic action
for electric and magnetic NS fluxes has not been worked out previously
and we fill this gap here. Furthermore, we need this
action in order to facilitate the comparison with
the mirror version derived in section \ref{LLEA}
for type IIA compactified on the manifolds $\hat Y$.
Since the derivation of this action closely follows
the existing literature \cite{BGHL,JM,TV,GD,LM2}
 we only highlight the aspects which are relevant for our
analysis.

The ten-dimensional bosonic spectrum of type IIB supergravity consists of the
metric $\hat{g}$, an antisymmetric tensor field $\hat{B_2}$ and the dilaton
$\hat{\phi}$ in the NS-NS sector and an  axion $\hat{l}$,  a 2-form
$\hat{C_2}$,  and a 4-form $\hat{A_4}$ with self-dual field strength
$* \hat F_5 = \hat F_5$, in the
RR sector. No local covariant action can be written for this theory in 10
dimensions due to the self-duality of $\hat F_5$.
Instead one uses the action \cite{JP}
\begin{eqnarray}
  \label{action}
  S_{IIB}^{(10)} &=& \int e^{-2\hat\phi}\left(-\frac{1}{2} R *\! {\bf 1} + 2
    \dd \hat\phi\wg 
    * \dd \hat \phi - \frac{1}{4} \dd \hat B_2\wg * \dd \hat
    B_2\right)\nonumber\\[2mm] 
  & & - \frac{1}{2} \int \left( \dd l\wg * \dd l+(\dd \hat C_2-l \dd \hat
    B_2)\wg 
    *(\dd \hat C_2-l \dd \hat B_2) + \frac12 \hat F_5 \wg * \hat
    F_5\right)\nonumber\\[2mm] 
  &&-\frac12\int\hat A_4\wg \dd \hat B_2\wg \dd\hat C_2 ,
\end{eqnarray}
where
\begin{eqnarray}
  \hat F_5 & = & \dd \hat A_4 - \dd \hat B_2 \wg \hat C_2\ ,
  \label{F5}
\end{eqnarray}
and imposes the self-duality of $\hat F_5$  separately.

The compactification proceeds as usual by expanding the ten-dimensional
quantities
in terms of harmonic forms on the \CY{} manifold
\begin{eqnarray}
  \label{fexpB}
  \hat B_2 & = &  B_2 +  b^i \wg \ox_i \ ,\qquad i = 1,\ldots,h^{(1,1)}\ , \\
  \hat C_2 & = & C_2 + c^i \wg \ox_i\ , \nn \\
  \hat A_4 & = & D_2^i \wg \ox_i + \rho_i \wg \tilde \ox^i + V^A \wg \ax_A
  - U_A \wg \bx^A\ ,\qquad A = 1,\ldots,h^{(1,2)} \ , \nn
\end{eqnarray}
where $B_2,C_2,D_2^i$ are two-forms, $V^A,U_A$ are one-forms
and $b^i,c^i,\rho_i$ are scalar fields in $D=4$. The
$\ox_i$ form a basis for the harmonic $(1,1)$-forms
and $(\ax_A\, ,\, \bx^A)$ form a basis for the harmonic
3-forms as introduced in the previous section.
The self-duality of  $\hat F_5$ implies that only
half of the fields appearing in the expansion of $\hat A_4$ in (\ref{fexpB})
are independent. The four-dimensional spectrum  consists of a
gravitational multiplet
$(g_{\mu \nu}, V_\mu^0)$, a double tensor multiplet $(B_2, C_2, \phi, l)$,
$h^{(1,1)}$ tensor multiplets $(D_2^i, v^i, b^i, c^i)$ and $h^{(1,2)}$ vector
multiplets $(V^a, z^a)$.  The $v^i$ represent the K\"ahler class moduli while
the $z^a$ are the complex structure moduli as introduced in \ref{IIAcomp}.
In Calabi--Yau compactifications without fluxes
all these fields are massless and the
tensor and double tensor multiplets can be dualized to $h^{(1,1)} +1$
hypermultiplets.

Turning on NS fluxes amounts to a modification of $H_3$  according to
\begin{equation}
  \label{conv2}
  \dd \hat B_2 = \dd B_2 + \dd b^i \wg \ox_i +\p^A\ax_A - \q_A \bx^A\ .
\end{equation}
After  taking into account the
self-duality of $F_5$ one arrives at the following action
\begin{eqnarray}
  S_{IIB}^{(4)} &=& \int - \frac{1}{2} R *\! {\bf 1} - g_{ab} \dd z^a \wg
  * \dd \bar{z}^{b} - g_{ij} \dd t^i \wg * \dd \bar{t}^j - \dd \phi \wg * \dd
  \phi \nonumber\\[2mm]
  && -\frac{1}{4} e^{-4\phi} \dd B_2 \wg * \dd B_2 - \frac{1}{2} e^{-2\phi} \cK
  \left( \dd C_2 - l \dd B_2 \right) \wg *\left( \dd C_2 - l \dd B_2 \right)
  \nonumber\\[2mm]
  && - \frac{1}{2} \cK e^{2\phi} \dd l \wg * \dd l - 2 \cK e^{2\phi} g_{ij}
  \left(\dd c^i - l \dd b^i \right)\wg * \left( \dd c^j - l \dd b^j
  \right)\nonumber\\[2mm] 
  && - \frac{e^{2\phi}}{2 \cK} g^{-1\,ij} \left( \dd \rho_i - \frac{1}{2}
    \cK_{ikl} c^k \dd b^l \right) \wg *\left( \dd \rho_j - \frac{1}{2}
    \cK_{jmn} c^m \dd b^n \right) \nonumber \\[2mm]
  && + 2 \left( \dd b^i \wg C_2 + c^i \dd B_2 \right)\wg
  \left( \dd \rho_i - \frac{1}{2} \cK_{ijk} c^j \dd b^k \right) + \frac{1}{2}
  \cK_{ijk} c^i c^j dB_2 \wg \dd b^k \nonumber \\[2mm]
  && + \frac{1}{2} \RE\cM_{AB} \tilde{F}^A \wg \tilde{F}^B + \frac{1}{2} \IM
  \cM_{AB} \tilde{F}^A \wg * \tilde{F}^B + \frac{1}{2} \q_A \left( F^A +
    \tilde{F}^A \right) \wg C_2\nn \\[2mm]
  && + \frac{1}{2} e^{4\phi} \left(l^2 + \frac{e^{-2\phi}}{2 \cK }\right)
  \left(\q - \cM \p \right)_A \IM \cM^{-1AB} \left(\q - \bar{\cM}
  \p\right)_B *\! {\bf 1} \ ,\label{action1}
\end{eqnarray}
where $\tilde F^A = F^A-\p^AC_2$ and the metrics $g_{ab}, g_{ij}$
as well as the other scalar dependent couplings have been defined in the
previous appendix.
Due to the appearance of $\tilde F^A$ in (\ref{action1})
 the RR 2-form $C_2$ is
massive. It was shown \cite{LM2} that in the case of only RR fluxes
the NS 2-form $B_2$ acquired a mass. Due to the $SL(2, \bf{R})$ symmetry
of the ten-dimensional type IIB effective action which rotates the two 2-forms into one another this is in
agreement with the result found here that when NS fluxes are present the RR
2-form $C_2$ becomes massive.

In most parts of this  paper we choose to consider $\p^A=0$
and in this case  all 2-forms are
massless and can be dualized to scalars.\footnote{In fact the massive
RR two-form is one of the technical reasons that the construction
of the mirror symmetric type IIA effective action is more involved.
We will come back to this issue in a separate publication \cite{GLMW}.}
After redefining these scalars  appropriately \cite{BGHL}
the sigma model metric for the hypermultiplets can be brought to the standard
quaternionic form of ref.~\cite{FeS}. In this basis the action reads
\begin{eqnarray}
  S_{IIB}^{(4)} & = & \int -\frac{1}{2}R * \! {\mathbf 1} - g_{a b} \dd z^a \wg
  *\dd \bar{z}^{b} - h_{uv} D q^u \wg * D q ^v - V_{IIB} * \! {\mathbf 1}
  \nonumber\\ 
  && + \frac{1}{2} \RE \cM_{AB} F^A \wg F^B + \frac{1}{2} \IM \cM_{AB}F^A \wg
  * F^B \ ,
  \label{action3}
\end{eqnarray}
where the quaternionic metric is given by
\begin{eqnarray}
h_{uv} D q^u \wg * D q ^v &=&
g_{ij} \dd t^i \wg * \dd \bar{t}^j + \dd \phi \wg * \dd \phi \\[2mm]
  && - \frac{1}{2} e^{2\phi} \IM \cN^{-1 \, IJ} \left(D \tilde{\xi}_I +
    \cN_{IK} D \xi^K \right) \wg * \left( D \tilde{\xi}_J + \bar{\cN}_{JL} D
    \xi^L \right) \nonumber\\[2mm]
  && + \frac{1}{4} e^{4\phi} \left(D a + (\xi^I D \tilde{\xi}_I -
    \tilde{\xi}_I D \xi^I) \right) \wg * \left(D a + (\xi^I D
    \tilde{\xi}_I - \tilde{\xi}_I D\xi^I) \right) \ , \nonumber
\end{eqnarray}
while the potential reads
\begin{equation}
  V_{IIB} = -\frac{1}{2}\, e^{4 \phi} \Big(l^2 + \frac{e^{-2\phi}}{2 \cK}
  \Big)\, \q_A  \left[(\IM \cM)^{-1}\right]^{AB}\q_B .
  \label{potIIB}
\end{equation}
The presence of the electric fluxes has gauged some of the isometries of the
hyperscalars as can be seen from the covariant derivatives
\begin{equation}
 D a  =  \dd a - \xi^0 \q_A V^A\ ,\qquad
  D\tilde{\xi}_0 =  \dd \tilde{\xi}_0 - \q_A V^A \ , \qquad
D\tilde{\xi}_i =  \dd \tilde{\xi}_i \ , \qquad
D \xi^I =  \dd \xi^I \ .
  \label{action5}
\end{equation}

\section{$G$-structures}
\label{acs}

In this section we assemble a few facts about $G$-structures as taken
from the mathematical literature where one also finds the proofs
omitted here. (See, for example,
\cite{FFS,salamonb,joyce,salamon,CS,yano,candelasTS}.) We concentrate on the
example of manifolds with $\SU(3)$-structure.

\subsection{Almost Hermitian manifolds}
\label{ahm}

Before discussing $G$-structures in general, let us recall the
definition of an almost Hermitian manifold. This allows us to introduce
some useful concepts, and, as we subsequently will see, provides us with
a classic example of a $G$-structure.  

A manifold of real dimension $2n$ is called \emph{almost complex} if
it admits a globally defined tensor field $J_m{}^n$ which obeys 
\begin{equation}\label{J2}
  J_m{}^p J_p{}^n = -\dx_m{}^n \ .
\end{equation}
A metric $g_{mn}$ on such a manifold is called Hermitian if it satisfies
\begin{equation}\label{hm}
  J_m{}^p J_n{}^r g_{pr} = g_{mn}\ .
\end{equation}
An almost complex  manifold endowed with a Hermitian metric is called
an \emph{almost Hermitian manifold}. The relation (\ref{hm}) implies that
$J_{mn} = J_m{}^p g_{pn}$ is a non-degenerate 2-form which is called
\emph{the fundamental form}. 

On any even-dimensional manifold one can locally introduce complex
coordinates. However, complex manifolds have to satisfy in addition
that, first, the introduction of complex coordinates on different
patches is consistent, and second that the transition functions
between different patches are holomorphic functions of the complex
coordinates. The first condition corresponds to the existence of an
almost complex structure. The second condition is an integrability
condition, implying that there are coordinations such that the almost
complex structure takes the form 
\begin{equation}\label{diagJ}
  J = \left(
  \begin{array}{cc}
    i \bf{1}_{n\times n} & 0 \\
    0 & -i \bf{1}_{n\times n}
  \end{array}\right) \ .
\end{equation}
The integrability condition is satisfied if and only if the Nijenhuis
tensor $N_{mn}{}^p$ vanishes. It is defined as
\begin{equation}
\label{Ntens}
\begin{split}
   N_{mn}{}^p &= J_m{}^q \left (\partial_q J_n{}^p 
      - \partial_n J_q{}^p \right) 
      - J_n{}^q \left (\partial_q J_m{}^p - \partial_m J_q{}^p \right)
   \\ 
      &= J_m{}^q \left (\nabla_q J_n{}^p - \nabla_n J_q{}^p \right)
      - J_n{}^q \left (\nabla_q J_m{}^p - \nabla_m J_q{}^p \right) \ ,
\end{split}
\end{equation}
where $\nabla$ denotes the covariant derivative with respect to the
Levi--Civita connection. 

One can also consider an even stronger condition where
$\nabla_mJ_{np}=0$. This implies $N_{mn}{}^p=0$ but in addition that
$\dd J=0$ and means we have a {\em K\"ahler manifold}. In particular,
it implies that the holonomy of the Levi--Civita connection $\nabla$
is $U(n)$. 

Even if there is no coordinate system where it can be put in the
form~\eqref{diagJ}, any almost complex structure obeying (\ref{J2})
has eigenvalues $\pm i$. Thus even for non-integrable almost complex
structures one can define the projection operators 
\begin{equation}\label{Pdef}
  (P^\pm)_m{}^n = \frac12(\dx_m^n \mp i J_m{}^n)\ ,
\end{equation}
which project onto the two eigenspaces, and satisfy
\begin{equation}\label{Pprop}
P^\pm P^\pm = P^\pm \ , \qquad P^+ P^- =0\ .
\end{equation}
On an almost complex manifold one can define $(p,q)$ projected components
$\ox^{p,q}$ of a real $(p+q)$-form $\ox^{p+q}$
by using (\ref{Pdef})
\begin{equation}
  \ox^{p,q}_{m_1\ldots m_{p+q}} = 
    (P^+)_{m_1}{}^{n_1} \ldots (P^+)_{m_p}{}^{n_p}
    (P^-)_{m_{p+1}}{}^{n_{p+1}} \ldots (P^-)_{m_{p+q}}{}^{n_{p+q}}
\ox_{n_1  \ldots n_{p+q}}^{p+q} \ .
\end{equation}
Furthermore, a real $(p+q)$-form is of the type $(p,q)$ if
it satisfies
\begin{equation}
  \ox_{m_1 \ldots m_p n_1 \ldots n_q} = 
     (P^+)_{m_1}{}^{r_1} \ldots (P^+)_{m_q}{}^{r_p} 
     (P^-)_{n_1}{}^{s_1} \ldots (P^-)_{n_q}{}^{s_q} \ox_{r_1
  \ldots r_p s_1 \ldots s_q} \ .
\end{equation}

In analogy with complex manifolds  we denote  the projections
on the subspace of eigenvalue $+i$ with an
unbarred  index $\ax$ and the projection on the subspace of
eigenvalue $-i$ with a barred index $\ab$.
For example the hermitian metric of an almost Hermitian manifold is of type
$(1,1)$ and has one barred and one unbarred index.
Thus, raising and lowering indices using this hermitian metric
converts holomorphic indices into anti-holomorphic ones and vice versa.
Moreover the contraction of a
holomorphic and an anti-holomorphic index vanishes, i.e.\ given $V_m$ which is
of type $(1,0)$ and $W^n$ which is of type $(0,1)$, the product $V_m W^m$
is zero. Similarly, on an almost hermitian manifold of real dimension $2n$
forms of type $(p,0)$ vanish for $p>n$.
Finally, derivatives of $(p,q)$-forms pick up extra pieces compared to
complex manifolds precisely because $J$ is not constant. One finds
\cite{candelasTS} 
\begin{equation}
\label{ddecomp}
  \dd \ox^{(p,q)} = (\dd \ox)^{(p-1, q+2)} + (\dd \ox)^{(p, q+1)} + (\dd
  \ox)^{(p+1,q)} + (\dd \ox)^{(p+2, q-1)} \ .
\end{equation}

\subsection{$G$-structures and $G$-invariant tensors}
\label{Gstruc}

An orthonormal frame on a $d$-dimensional Riemannian manifold $M$ is
given by a basis of vectors $e_i$, with $i=1,\dots,d$,
satisfying $e_i^me_j^ng_{mn}=\delta_{ij}$. The set of all orthonormal
frames is known as the frame bundle. In general, the structure group
of the frame bundle is the group of rotations $O(d)$ (or $\SO(d)$ is
$M$ is orientable). The manifold has a $G$-structure if the 
structure group of the frame bundle is not completely general but can
be reduced to $G\subset O(d)$. For example, in the case of an almost
Hermitian manifold of dimension $d=2n$, in turns out one can always
introduce a complex frame and as a result the structure group reduces
to $U(n)$. 

An alternative and sometimes more convenient way to define
$G$-structures is via $G$-invariant tensors, or, if $M$ is spin,
$G$-invariant spinors. A non-vanishing, globally defined tensor or
spinor $\xi$ is $G$-invariant if it is invariant under $G\subset
O(d)$ rotations of the orthonormal frame. In the case of almost
Hermitian structure, the two-form $J$ is an $U(n)$-invariant tensor.
Since the invariant tensor $\xi$ is globally defined, by considering
the set of frames for which $\xi$ takes the same fixed form, one can
see that the structure group of the frame bundle must then reduce to
$G$ (or a subgroup of $G$). Thus the existence of $\xi$ implies we
have a $G$-structure. Typically, the converse is also true. Recall
that, relative to an orthonormal frame, tensors of a given type form
the vector space for a given representation of $O(d)$ (or $\Spin(d)$
for spinors). If the structure group of the frame bundle is reduced to
$G\subset O(d)$, this representation can be decomposed into
irreducible representations of $G$. In the case of almost complex
manifolds, this corresponds to the decomposition under the $P^\pm$
projections~\eqref{Pdef}.  Typically there will be some tensor or
spinor that will have a component in this decomposition which is
invariant under $G$. The corresponding vector bundle of this component
must be trivial, and thus will admit a globally defined non-vanishing
section $\xi$. In other words, we have a globally defined
non-vanishing $G$-invariant tensor or spinor.

To see this in more detail in the almost complex structure example,
recall that we had a globally defined fundamental two-form $J$. Let us
specialize for definiteness to a six-manifold, though the argument is
quite general. Two-forms are in the adjoint representation $\rep{15}$
of $\SO(6)$ which decomposes under $U(3)$ as 
\begin{equation}
   \rep{15} = \rep{1} + \rep{8} + (\rep{3} + \bar{\rep{3}})\  .
\end{equation}
There is indeed a singlet in the decomposition and so given a
$U(3)$-structure we necessarily have a globally defined invariant
two-form, which is precisely the fundamental two-form $J$. Conversely,
given a metric and a non-degenerate two-form $J$, we have an almost
Hermitian manifold and consequently an $U(3)$-structure.   

In this paper we are interested in $\SU(3)$-structure. In this case we
find two invariant tensors. First we have the fundamental form $J$ as
above. In addition, we find an invariant complex
three-form $\Omega$. Three-forms are in the $\rep{20}$ representation
of $\SO(6)$, giving two singlets in the decomposition under
$\SU(3)$,  
\begin{equation}
\begin{aligned}
   \rep{15} &= \rep{1} + \rep{8} + \rep{3} + \bar{\rep{3}}
       \quad \Rightarrow \quad J \  , \\
   \rep{20} &= \rep{1} + \rep{1} + \rep{3} + \bar{\rep{3}} 
       + \rep{6} + \bar{\rep{6}} 
       \quad \Rightarrow \quad \Omega = \Omega^+ +\ii\Omega^- \ .
 \end{aligned}
 \end{equation}
In addition, since there is no singlet in the decomposition of a
five-form, one finds that 
\begin{equation}\label{JOcond}
   J \wedge \Omega = 0 \ . 
\end{equation}
Similarly, a six-form is a singlet of $\SU(3)$, so we also must have
that $J\wedge J\wedge J$ is proportional to
$\Omega\wedge\bar\Omega$. The usual convention is to set
\begin{equation}
\label{JOcond2}
   J\wedge J \wedge J
      = \frac{3\ii}{4}\, \Omega \wedge \bar{\Omega} \ , \\
\end{equation}
Conversely, a non-degenerate $J$ and $\Omega$
satisfying~\eqref{JOcond} and~\eqref{JOcond2} implies that $M$ has 
$\SU(3)$-structure. Note that, unlike the $U(n)$ case, the
metric need not be specified in addition; the existence of $J$ and
$\Omega$ is sufficient~\cite{H}. Essentially this is because, without
the presence of a metric, $\Omega$ defines an almost complex
structure, and $J$ an almost symplectic structure. Treating $J$ as the
fundamental form, it is then a familiar result on almost Hermitian
manifolds that the existence of an almost complex structure and a
fundamental form allow one to construct a Hermitian metric.  

We can similarly ask what happens to spinors for a structure group $SU(3)$.
In this case we have the isomorphism $\Spin(6)\cong\SU(4)$  and
the four-dimensional spinor representation decomposes as 
\begin{equation}
   \rep{4}=\rep{1}+\rep{3} \quad \Rightarrow \quad \eta \ .
\end{equation}
We find one singlet in the decomposition, implying the existence of a
globally defined invariant spinor $\eta$. Again, the converse is also
true. A metric and a globally defined spinor $\eta$ implies that $M$
has $\SU(3)$-structure.

\subsection{Intrinsic torsion}
\label{app:IT}

One would like to have some classification of $G$-structures. In
particular, one would like a generalization of the notion of a
K\"ahler manifold where the holonomy of the Levi--Civita connection
reduces to $U(n)$. Such a classification exists in terms of the {\em
  intrinsic torsion}. Let us start by recalling the definition of
torsion and contorsion on a Riemannian manifold $(M,g)$.  

Given any metric compatible connection $\nabla'$ on $(M,g)$, i.e.
one satisfying $\nabla'_mg_{np}=0$, one can define the Riemann
curvature tensor and the torsion tensor as follows
\begin{equation}
\label{RT}
  [\nabla'_m, \nabla'_n] V_p = 
     - R_{mnp}{}^q V_q - 2 T_{mn}{}^r \nabla'_r V_p \ ,
\end{equation}
where $V$ is an arbitrary vector field. The Levi-Civita connection is
the unique torsionless connection compatible with the metric and is
given by the usual expression in terms of Christoffel symbols
$\Gx_{mn}{}^p = \Gx_{nm}{}^p$. Let us denote by $\nabla$ the 
covariant derivative with respect to the Levi-Civita connection while a
connection with torsion is denoted by $\nabla^{(T)}$.
Any metric compatible connection can be written in terms of the
Levi-Civita connection 
\begin{equation}
  \label{cont}
  \nabla^{(T)} = \nabla + \CT \ ,
\end{equation}
where $\CT_{mn}{}^p$ is the contorsion tensor. Metric compatibility implies
\begin{equation}\label{Kprop}
  \CT_{mnp} = - \CT_{mpn}\ , \quad \textrm{where} \quad
\CT_{mnp} = \CT_{mn}{}^r g_{rp} \ .
\end{equation}
Inserting (\ref{Kprop}) into (\ref{RT}) one finds a one-to-one
correspondence between the torsion and the contorsion 
\begin{equation}
\begin{aligned}
  \label{T-K}
  T_{mn}{}^p &= 
    \frac12 (\CT_{mn}{}^p - \CT_{nm}{}^p) \equiv \CT_{[mn]}{}^p \ , \\
  \CT_{mnp} &=  T_{mnp} + T_{pmn} + T_{pnm} \ .
\end{aligned}
\end{equation}
These relations tell us that given a torsion tensor $T$ there exist a
unique connection $\nabla^{(T)}$ whose torsion is precisely $T$.

Now suppose $M$ has a $G$-structure. In general the Levi-Civita
connection does not preserve the $G$-invariant tensors (or
spinor) $\xi$. In other words, $\nabla\xi\neq 0$. However, one can
show~\cite{joyce}, that there always exist some other connection $\nabla^{(T)}$
which is compatible with the $G$ structure so that 
\begin{equation}
   \nabla^{(T)}\xi = 0\ . 
\end{equation}
Thus for instance, on an almost Hermitian manifold one can always find
$\nabla^{(T)}$ such that $\nabla^{(T)} J=0$. On a manifold with
$\SU(3)$-structure, it means we can always find $\nabla^{(T)}$ such
that both $\nabla^{(T)} J=0$ and $\nabla^{(T)}\Omega=0$. Since
the existence of $\SU(3)$-structure is also equivalent to the
existence of an invariant spinor $\eta$, this is equivalent to the
condition $\nabla^{(T)}\eta=0$. 

Let $\CT$ be the contorsion tensor corresponding to
$\nabla^{(T)}$. From the symmetries~\eqref{Kprop}, we see that $\CT$
is an element of $\Lambda^1\otimes\Lambda^2$ where $\Lambda^n$ is the
space of $n$-forms. Alternatively, since $\Lambda^2\cong\so(d)$, it is
more natural to think of $\CT_{mn}{}^p$ as one-form with values in the
Lie-algebra $\so(d)$ that is $\Lambda^1\otimes\so(d)$. Given the
existence of a $G$-structure, we can decompose $\so(d)$ into a
part in the Lie algebra $g$ of $G\subset\SO(d)$ and an orthogonal
piece $g^\perp=\so(d)/g$. The contorsion $\CT$ splits according into 
\begin{equation}
   \CT = \CT^0 + \CT^g \ ,
\end{equation}
where $\CT^0$ is the part in $\Lambda^1\otimes g^\perp$. Since an
invariant tensor (or spinor) $\xi$ is fixed under $G$ rotations, the
action of $g$ on $\xi$ vanishes and we have, by definition, 
\begin{equation}
\label{ICTdef}
   \nabla^{(T)}\xi = \left(\nabla + \CT^0 + \CT^g\right)\xi
       = \left(\nabla + \CT^0\right)\xi = 0 \ .
\end{equation}
Thus, any two $G$-compatible connections must differ by a piece
proportional to $\CT^g$  and they have a common term $\CT^0$ in
$\Lambda^1\otimes g^\perp$ called the ``intrinsic contorsion''. Recall
that there is an isomorphism~\eqref{T-K} between $\CT$ and $T$. It is
more conventional in the mathematics literature to define the
corresponding torsion 
\begin{equation}
\label{IT}
   T^0_{mn}{}^p = \CT^0_{[mn]}{}^p \in \Lambda^1\otimes g^\perp \ ,
\end{equation}
known as the {\em intrinsic torsion}. 

From the relation~\eqref{ICTdef} it is clear that the intrinsic
contorsion, or equivalently torsion, is independent of the choice of
$G$-compatible connection. Basically it is a measure of the degree to
which $\nabla\xi$ fails to vanish and as such is a measure solely of
the $G$-structure itself. Furthermore, one can decompose $\CT^0$ into
irreducible $G$ representations. This provides a classification of
$G$-structures in terms of which representations appear in the
decomposition. In particular, in the special case where $\CT^0$
vanishes so that $\nabla\xi=0$, one says that the structure is
``torsion-free''. For an almost Hermitian structure this is
equivalent to requiring that the manifold is complex and K\"ahler. In
particular, it implies that the holonomy of the Levi--Civita
connection is contained in $G$. 

Let us consider the decomposition of $T^0$ in the case of
$SU(3)$-structure. The relevant representations are 
\begin{equation}
\Lambda^1 \sim \rep 3\oplus \rep{\bar 3}\ ,\qquad
g \sim \rep 8\ , \qquad g^\perp \sim \rep 1 \oplus \rep 3\oplus \rep{\bar 3}\ .
\end{equation}
Thus the intrinsic torsion, which is an element of
$\Lambda^1\otimes\su(3)^\perp$, can be decomposed into the following
$SU(3)$ representations
\begin{equation}
\label{ITdecomp}
\begin{split}
   \Lambda^1 \otimes \su(3)^\perp &= 
      (\rep 3\oplus \rep{\bar 3}) \otimes 
         (\rep 1 \oplus \rep 3\oplus\rep{\bar 3 }) \\
      &= (\rep 1 \oplus \rep 1) \oplus (\rep 8 \oplus \rep 8) 
         \oplus (\rep 6 \oplus \rep{\bar  6}) 
         \oplus (\rep 3 \oplus \rep{\bar  3}) 
         \oplus (\rep 3 \oplus \rep{\bar  3})' \ .
\end{split}
\end{equation}
The terms in parentheses on the second line correspond precisely to
the five classes $\W_1,\ldots,\W_5$ presented in table~\ref{tabW}. We
label the component of $T^0$ in each class by $T_1,\dots,T_5$.

In the case of $\SU(3)$-structure, each component $T_i$ can be
related to a particular component in the $\SU(3)$ decomposition of
$\dd J$ and  $\dd\Omega$. From~\eqref{ICTdef}, we have 
\begin{equation}
\label{dJdO}
\begin{aligned}
   \dd J_{mnp} &= 6 T^0_{[mn}{}^r J_{r|p]} \ , \\
   \dd \Ox_{mnpq} &= 12 T^0_{[mn}{}^r \Ox_{r|pq]}\ .   
\end{aligned}
\end{equation}
Since $J$ and $\Omega$ are $\SU(3)$ singlets, $\dd J$ and $\dd\Omega$
are both elements of $\Lambda^1\otimes\su(3)^\perp$. Put another way,
the contractions with $J$ and $\Omega$ in~\eqref{dJdO} simply project
onto different $\SU(3)$ representations of $T^0$. We can see which
representations appear simply by decomposing the real three-form $\dd
J$ and complex four-form $\dd\Omega$ under $\SU(3)$. We have, 
\begin{equation}
\label{dJdecomp}
\begin{aligned}
   \dd J &= \big[(\dd J)^{3,0} + (\dd J)^{0,3} \big]
       + \big[(\dd J)^{2,1}_0 + (\dd J)^{1,2}_0 \big]
       + \big[(\dd J)^{1,0} + (\dd J)^{0,1} \big] \ , \\
   \rep{20} &= (\rep 1 \oplus \rep 1)
       \oplus (\rep 6 \oplus \rep{\bar  6}) 
       \oplus (\rep 3 \oplus \rep{\bar  3}) \ ,
\end{aligned}
\end{equation}
and
\begin{equation}
\label{dOdecomp}
\begin{aligned}
   \dd \Omega &= (\dd \Omega)^{3,1} + (\dd\Omega)^{2,2}_0  +
      (\dd\Omega)^{0,0} \ , \\  
   \rep{24} &= (\rep 3 \oplus \rep{\bar  3})' 
      \oplus (\rep 8 \oplus \rep 8) 
      \oplus (\rep 1 \oplus \rep 1) \ .
\end{aligned}
\end{equation}
The superscripts in the decomposition of $\dd J$ and $\dd\Omega$ refer 
to the $(p,q)$-type of the form. The $0$ subscript refers to the
irreducible $\SU(3)$ representation where the trace part, proportional
to $J^n$ has been removed. Thus in particular, the traceless parts $(\dd
J)^{2,1}_0$ and $(\dd\Omega)^{2,2}_0$ satisfy $J\wedge(\dd
J)^{2,1}_0=0$ and $J\wedge(\dd\Omega)^{2,2}_0=0$ respectively. The
trace parts on the other hand, have the form $(\dd
J)^{1,0}=\alpha\wedge J$ and $(\dd\Omega)^{0,0}=\beta J\wedge
J$, with $\alpha\sim *(J\wedge\dd J)$ and $\beta\sim
*(J\wedge\dd\Omega)$ respectively. 
Note that a generic complex four-form has 30 components. However,
since $\Omega$ is a $(3,0)$-form, from~\eqref{ddecomp} we see that
$\dd\Omega$ has no $(1,3)$ part, and so only has 24
components. Comparing~\eqref{dJdecomp} and~\eqref{dOdecomp}
with~\eqref{ITdecomp} we see that 
\begin{equation}
   \dd J \in \W_1 \oplus \W_3 \oplus \W_4 \ , \quad
   \dd\Omega \in \W_1 \oplus \W_2 \oplus \W_5 \ ,
\end{equation}
and as advertised, $\dd J$ and $\dd\Omega$ together include all the
components $T_i$. Explicit expressions for some of these relations are
given above in~\eqref{cxT12} and \eqref{T3}. Note that the singlet
component $T_1$ can be expressed either in terms of $(\dd J)^{0,3}$,
corresponding to $\Omega\wedge\dd J$ or in terms of
$(\dd\Omega)^{0,0}$ corresponding to $J\wedge\dd\Omega$. This is
simply a result of the relation~\eqref{JOcond} which implies that
$\Omega\wedge\dd J=J\wedge\dd\Omega$.


\section{The Ricci scalar of half-flat manifolds}
\label{Rhf}

The simplest way to derive the Ricci scalar for the manifold considered in
section \ref{Yhat} is by using the integrability condition one can derive from
the Killing spinor equation (\ref{torKS}).
\begin{equation}
  \label{intKS1}
  R^{(T)}_{mnpq} \Gx^{pq} \eta = 0,
\end{equation}
where the Riemann tensor of the connection with torsion is given by (\ref{RTexp})
\begin{equation}
  \label{RT1}
  R_{mnpq}^{(T)} = R(\Gx)_{mnpq} + \nabla_m \CT_{npq} - \nabla_n \CT_{mpq} -
  \CT_{mp}{}^r \CT_{nrq} + \CT_{np}{}^r \CT_{mrq} \ .
\end{equation}
Here $R(\Gx)_{mnpq}$ represents the usual Riemann tensor for the Levi-Civita
connection and the covariant derivatives are again with respect to the
Levi-Civita connection.
For definiteness we choose the solution of the Killing spinor equation
(\ref{torKS}) to be a Majorana spinor.\footnote{The results are independent of
the choice of the spinor, but the derivations may be more involved.}
Multiplying (\ref{intKS1}) by $\Gx^n$ and summing over $n$ one obtains
\begin{equation}
  \label{intKS2}
  R^{(T)}_{mnpq} \Gx^{npq} \eta - 2 R^{(T)}_{mn} \Gx^{n} \eta = 0\ .
\end{equation}
Contracting from the left with $\eta^\dagger \Gx^m$ and using the conventions
for the Majorana spinors (\ref{asym}) one derives
\begin{equation}
  \label{intKS3}
  2 R^{(T)} = R^{(T)}_{mnpq} \eta^\dagger \Gx^{mnpq} \eta\ .
\end{equation}
where $R^{(T)}$ represents the Ricci scalar which can be defined from the
Riemann tensor (\ref{RT1}). 
Expressing $ R^{(T)}_{mnpq}$ in terms of $R(\Gx)_{mnpq}$ from (\ref{RT1}),
using the Bianchi identity $R(\Gx)_{m[npq]} = 0$ and the fact that the
contorsion is traceless $\CT_{mn}{}^m = \CT^m{}_{mn} = 0$ which holds for half
flat manifolds one
can derive the formula for the Ricci scalar of the Levi-Civita connection
\begin{equation}
  \label{Rapp}
  R = - \CT_{mnp} \CT^{npm} - \frac{1}{2} \epsilon^{mnpqrs} (\nabla_m \CT_{npq} -
  \CT_{mp}{}^l \CT_{nlq}) J_{rs} \ .
\end{equation}

In order to simplify the formulas we evaluate (\ref{Rapp})
term by term. The strategy will be to express first the contorsion $\CT$ in
terms of the torsion $T$ (\ref{T-K}) and then go to complex indices splitting
the torsion in its component parts $T_{1\oplus2}$ and $T_3$ which are of
definite type with respect to the almost complex structure $J$.

The first term can be written as
\begin{equation}
  \label{A1}
  A \equiv - \CT_{mnp} \CT^{npm} = - (T_{mnp} + T_{pmn} + T_{pnm}) T^{npm} = T_{mnp}
  T^{mnp} - 2 T_{mnp} T^{npm}.
\end{equation}
Using (\ref{su3Nt}) and (\ref{T3}) one sees that the first two indices of
$T$ are of the same type and thus one has
\begin{equation}
  \label{Af}
  A = (T_{1\oplus 2})_{\ax \bx \cx} (T_{1\oplus 2})^{\ax \bx \cx} - 2
  (T_{1\oplus 2})_{\ax \bx \cx} (T_{1\oplus 2})^{\bx \cx \ax}  + (T_3)_{\ax
  \bx \cb} (T_3)^{\ax \bx \cb} + c.c. \ ,
\end{equation}
where $c.c.$ denotes complex conjugation.

The second term can be computed if one takes into account that the
four-dimensional effective action appears after one integrates the
ten-dimensional 
action over the internal space, in this case  $\hat Y$.
Thus the second term in (\ref{Rapp}) can be integrated by parts to
give\footnote{
Strictly speaking in 10 dimensions the Ricci scalar comes multiplied with a
dilaton factor (\ref{SIIA}). However in all what we are doing we consider that
the dilaton is constant over the internal space so it still make sense to
speak about integration by parts without introducing additional factors with
derivatives of the dilaton. }
\begin{equation}
  \label{B1}
  B \equiv - \frac12 \epsilon^{mnpqrs} (\nabla_m \CT_{npq}) J_{rs} \sim
  \frac12 \epsilon^{mnpqrs} \CT_{npq} \nabla_m J_{rs}.
\end{equation}
Using (\ref{torJO}) and
(\ref{T-K}) we obtain after going to complex indices
\begin{eqnarray}
  \label{B2}
  B & =& - \epsilon^{mnpqrs} T_{mnp} T_{qr}{}^t J_{ts} \\
  &=& -\epsilon^{\ax \bx \cx
  \ab \bb \cb} (T_{1\oplus 2})_{\ax \bx \cx} (T_{1\oplus 2})_{\ab \bb}{}^\dx
  J_{\dx \cb} -\epsilon^{\ax \bx \cb \ab \bb \cx} (T_3)_{\ax \bx \cb}
  (T_3)_{\ab \bb}{}^{\bar \dx} J_{\bar \dx \cx} + c.c. \ . \nonumber
\end{eqnarray}
The six-dimensional $\epsilon$ symbol splits as
\begin{equation}
  \epsilon^{\ax \bx \cx \ab \bb \cb} = -i \epsilon^{\ax \bx \cx} \epsilon^{\ab
  \bb \cb} \ , 
\end{equation}
and after some algebra involving the three-dimensional $\epsilon$ symbol one
finds 
\begin{equation}
  \label{Bf}
  B = -2 (T_{1\oplus 2})_{\ax \bx \cx} (T_{1\oplus 2})^{\ax \bx \cx} -
  4(T_{1\oplus 2})_{\ax \bx \cx} (T_{1\oplus 2})^{\bx \cx \ax} - 2 (T_3)_{\ax
  \bx \cb} (T_3)^{\ax \bx \cb} + c.c. \ .
\end{equation}

In the same way one obtains for the last term
\begin{equation}
  \label{Cf}
  C \equiv \frac12 \epsilon^{mnpqrs} \CT_{mp}{}^t \CT_{ntq} J_{rs} = 2
  (T_{1\oplus 2})_{\ax \bx \cx} (T_{1\oplus 2})^{\ax \bx \cx} + 2  (T_3)_{\ax
  \bx \cb} (T_3)^{\ax \bx \cb} + c.c. \ .
\end{equation}
Collecting the results from (\ref{Af}), (\ref{Bf}) and (\ref{Cf}) the formula
for the Ricci scalar (\ref{Rapp}) becomes
\begin{equation}
  \label{Rint}
  R = (T_{1\oplus 2})_{\ax \bx \cx} (T_{1\oplus 2})^{\ax \bx \cx} - 6
  (T_{1\oplus 2})_{\ax \bx \cx} (T_{1\oplus 2})^{\bx \cx \ax}  + (T_3)_{\ax
  \bx \cb} (T_3)^{\ax \bx \cb} + c.c. \ .
\end{equation}
The first two terms in the above expression can be straightforwardly computed
using (\ref{su3Nt}), (\ref{F4}).  After a little algebra we find
\begin{eqnarray}
  \label{AB}
  (T_{1\oplus 2})_{\ax \bx \cx} (T_{1\oplus 2})^{\ax \bx \cx} & = & \frac{e_i e_j}{8 ||\Ox||^2}
  (\tox^i)_{\ax \bx \ab \bb} (\tox^j)^{\ax \bx \ab \bb}  \\
  (T_{1\oplus 2})_{\ax \bx \cx} (T_{1\oplus 2})^{\bx \cx \ax} & = & - \frac{e_i e_j}{8 ||\Ox||^2}
  (\tox^i)_{\ax \bx \ab \bb} (\tox^j)^{\ax \bx \ab \bb} + \frac{e_i e_j}{4
  ||\Ox||^2} (*\tox^i)_{\ax \bb} (*\tox^j)^{\ax \bb} + \frac{(e_i v^i)^2}{4
  ||\Ox ||^2 \cK^2} \ . \nn
\end{eqnarray}
In order to obtain the above expressions we have used (\ref{Jexp})
and  \cite{AS3}
\begin{equation}\label{D.15}
  (\tox^i)_{\ax \bx}{}^{\ax \bx} = \frac{2 v^i}{\cK} \ .
\end{equation}

Integrating (\ref{AB}) over $\hat Y$ we obtain
\begin{eqnarray}
  \label{ABf}
  \int_{\hat Y} (T_{1\oplus 2})_{\ax \bx \cx} (T_{1\oplus 2})^{\ax \bx \cx} & = &
  \frac{e_i e_j g^{ij}}{8 ||\Ox||^2 \cK}\ , \\
  \int_{\hat Y} (T_{1\oplus 2})_{\ax \bx \cx} (T_{1\oplus 2})^{\bx \cx \ax} & = &
  - \frac{e_i e_j g^{ij}}{16 ||\Ox||^2 \cK} + \frac{(e_i v^i)^2}{4 ||\Ox ||^2
  \cK} \ .\nn
\end{eqnarray}

Finally, we have to compute the third term in (\ref{Rint}).
For this we note that the expression \eqref{T3} for the $T_3$ component of
the intrinsic torsion can be written as
\begin{equation}
  \label{T3new}
  (\dd J)_{mnp} = 4 F (\Ox^-)_{mnp} + 6 (T_3)_{[mn}{}^r J_{|r|p]} \ .
\end{equation}
In order to evaluate this formula we need the expressions for $F$ and $\dd
J$ which correspond to the definition \eqref{F4}.
Using \eqref{D.15} one finds for $F$
\begin{equation}
  \label{TrF}
  F \equiv F_{\ax \bx}{}^{\ax \bx} = \frac{e_i v^i}{2 \cK ||\Ox||^2} \ .
\end{equation}
Taking the square of \eqref{T3new} we note that the terms on the RHS do
not mix as they carry indices of different types. Inserting \eqref{TrF}
and $\dd J$ of \eqref{eq:dJOe} we obtain
\begin{equation}
  \label{chi2}
  (e_i v^i)^2 \int_{\hat Y} \bx^0 \wg * \bx^0 = 2i \left(\frac{e_i
  v^i}{||\Ox||^2 \cK} \right)^2 \int_{\hat Y} \Ox \wg \bar \Ox + 2
\int_{\hat
  Y} (T_3)_{mnp} (T_3)^{mnp}\ .
\end{equation}
The integral which appears on the LHS is given by
\begin{equation}
  \label{imM}
  \int \bx^0 \wg * \bx^0 = - \left[(\IM \cM)^{-1} \right]^{00}=
  \frac{8}{||\Ox||^2 \cK} \ ,
\end{equation}
where the first equation follows from (\ref{star}) and (\ref{A-N})
while the second equation is less obvious.
The simplest way to see this is by using a mirror symmetry argument. 
We know that under mirror symmetry the
gauge couplings $\cM$ and $\cN$ are mapped into one another. This also means
that $(\IM \cM)^{-1}$ is mapped into $(\IM \cN)^{-1}$ and this matrix is given
in (\ref{ImN-1e}) for a Calabi--Yau space. From here one sees that the element
$\left[(\IM \cN)^{-1}\right]^{00}$ is just the inverse volume of the mirror
Calabi--Yau space. Using again mirror symmetry
and the fact that the K\"ahler potential of the K\"ahler moduli
(\ref{KpotK}) is mapped into the K\"ahler potential of the complex structure
moduli (\ref{Kpotcs}) we end up with the RHS
of the above equation.

Now we can write (\ref{chi2}) as
\begin{equation}
  \int_{\hat Y} (T_3)_{mnp} (T_3)^{mnp} = 3 \frac{(e_i v^i)^2}{||\Ox||^2 \cK} \ ,
\end{equation}
or in complex indices
\begin{equation}
   \label{C'f}
  \int_{\hat Y} (T_3)_{\ax \bx \cb} (T_3)^{\ax \bx \cb} = \frac32 \; \frac{(e_i
  v^i)^2}{||\Ox||^2 \cK} \ .
\end{equation}

Inserting (\ref{ABf}) and (\ref{C'f}) into (\ref{Rint}) and taking into
account that all the terms in (\ref{ABf}) and (\ref{C'f}) are explicitly real
such that the term `$c.c.$' in (\ref{Rint}) just introduces one more factor of
$2$ we obtain the final form of the Ricci scalar
\begin{equation}
  \label{Rfin}
  R = - \frac{1}{8} \, e_i e_j g^{ij} \left[(\IM \cM)^{-1} \right]^{00} \ ,
\end{equation}
where we have used again (\ref{imM}).

\vskip 1cm

\subsection*{Acknowledgments}

This work is supported by DFG -- The German Science Foundation,
GIF -- the German--Israeli Foundation for Scientific Research,
the European RTN Program HPRN-CT-2000-00148, the
DAAD -- the German Academic Exchange Service,
the CNRS -- the French Center
for National Scientific Research and the Royal Society, UK.

We have greatly benefited from conversations with R.~Donagi,
J.~Gauntlett, S.~Gukov, M.~Haack, A.~Huber-Klawitter,
C.~Hull, S.~Kachru, P.~Kaste, J.~Jost, D.~L\"ust,  M.~Reid, M.~Schulz, 
M.~Schwarz, C.~Vafa. We thank the authors of \cite{kachru} for communicating
their results prior to publication.

This work was initiated at the Isaac Newton Institute 
for Mathematical Sciences, Cambridge during the M-theory workshop 2002.
J.L.\ and D.W.\ thank the Newton Institute for the kind hospitality and   
R.\ Dijkgraaf, M.\ Douglas, J.P.\ Gauntlett 
and C.\ Hull for organizing a stimulating and fruitful workshop.
D.W. would also very much like to thank the Aspen Center for
Physics for hospitality during the completion of this work.

A.M.\ would like to thank the High Energy Physics group at Queen Mary,
London for hospitality during his visit, where part of this work was
done.


\providecommand{\href}[2]{#2}
\end{document}